\documentclass[10pt,journal,compsoc,twoside]{IEEEtran}

\usepackage[nocompress]{cite}

\usepackage[pdftex]{graphicx}
\graphicspath{{figure/}}

\renewcommand{\baselinestretch}{0.9}

\usepackage{tabularx}
\usepackage{url}

\usepackage[cmex10]{amsmath}
\usepackage{array}

\usepackage{rotating}
\usepackage{makecell}

\usepackage[caption=true,font=footnotesize,labelfont=sf,textfont=sf]{subfig}

\usepackage{color}
\usepackage{wrapfig}

\newcommand{\ssection}[1]{\noindent \textbf{#1} }

\usepackage{xspace} \newcommand*{\eg}{e.g.\@\xspace}
\newcommand*{\ie}{i.e.\@\xspace}

\newcommand{\Fref}[1]{Figure~\ref{#1}}

\usepackage[shortlabels]{enumitem}

\begin{document}

\title{Way to Go! Automatic Optimization of Wayfinding Design}

\author{
	Haikun Huang,
    Ni-Ching Lin,
    Lorenzo Barrett,
    Darian Springer,\\
    Hsueh-Cheng Wang,
    Marc Pomplun,
	Lap-Fai Yu,~\IEEEmembership{Member,~IEEE}%
\IEEEcompsocitemizethanks{
\IEEEcompsocthanksitem H.~Huang, L.~Barrett, D.~Springer, M.~Pomplun and L.-F.~Yu are with the University of Massachusetts Boston.
\IEEEcompsocthanksitem N.-C.~Lin and H.-C.~Wang are with the National Chiao Tung University.
%\IEEEcompsocthanksitem L.-F.~Yu is with the University of Massachusetts Boston. The work reported herein was done in part when he was at the University of California, Los 
%Angeles, and at the Singapore University of Technology and Design.
%\IEEEcompsocthanksitem S.-K.~Yeung is with the Singapore University of Technology and Design.
%\IEEEcompsocthanksitem D.~Terzopoulos is with the University of California, Los Angeles.
}% <-this % stops a space
\thanks{Manuscript received ?? ??, 2017; revised ?? ??, 2017.}}

% \author{
% 	Haikun Huang,~\IEEEmembership{Member,~IEEE,}
% 	Lap-Fai Yu,~\IEEEmembership{Member,~IEEE,}
%     Hsueh-Cheng Wang,~\IEEEmembership{Member,~IEEE,}
%     Marc Pomplun,~\IEEEmembership{Member,~IEEE,}% <-this % stops a space
% \IEEEcompsocitemizethanks{
% \IEEEcompsocthanksitem H.K.Huang is with the University of Massachusetts Boston. The work reported herein was done in part when he was at the University of 
%Massachusetts Boston.
% \IEEEcompsocthanksitem L.-F.~Yu is with the University of Massachusetts Boston.
% \IEEEcompsocthanksitem N.W is with the Institute of Electrical and Control Engineering National Chiao Tung University(NCTU)
% \IEEEcompsocthanksitem M.P is with the University of Massachusetts Boston.
% %\IEEEcompsocthanksitem L.-F.~Yu is with the University of Massachusetts Boston. The work reported herein was done in part when he was at the University of California, Los 
%Angeles, and at the Singapore University of Technology and Design.
% %\IEEEcompsocthanksitem S.-K.~Yeung is with the Singapore University of Technology and Design.
% %\IEEEcompsocthanksitem D.~Terzopoulos is with the University of California, Los Angeles.
% }% <-this % stops a space
% \thanks{Manuscript received ?? ??, 2016; revised ?? ??, 2016.}}

%\author{submission No.}

\markboth{Transactions on Visualization and Computer Graphics,~Vol.~??, No.~??, ??~??}%
{Huang \MakeLowercase{\textit{et al.}}: Way to Go! Automatic Optimization of Wayfinding Design}

\IEEEpubid{\makebox[\columnwidth]{\hfill 0000--0000/00/\$00.00~\copyright~2017 IEEE}%
\hspace{\columnsep}\makebox[\columnwidth]{Published by the IEEE Computer Society\hfill}}

\IEEEtitleabstractindextext{%

\begin{abstract}
\linespread{1.0}\selectfont
  Wayfinding signs play an important role in guiding users to navigate
  in a virtual environment and in helping pedestrians to find their
  ways in a real-world architectural site. Conventionally, the
  wayfinding design of a virtual environment is created manually, so
  as the wayfinding design of a real-world architectural site. The
  many possible navigation scenarios, as well as the interplay between
  signs and human navigation, can make the manual design process
  overwhelming and non-trivial. As a result, creating a wayfinding
  design for a typical layout can take months to several
  years~\cite{swd}. In this paper, we introduce the Way to Go!
  approach for automatically generating a wayfinding design for a
  given layout. The designer simply has to specify some navigation
  scenarios; our approach will automatically generate an optimized
  wayfinding design with signs properly placed considering human
  agents' visibility and possibility of making mistakes during a
  navigation. We demonstrate the effectiveness of our approach in
  generating wayfinding designs for different layouts such as a train
  station, a downtown and a canyon. We evaluate our results by
  comparing different wayfinding designs and show that our optimized
  wayfinding design can guide pedestrians to their destinations
  effectively and efficiently. Our approach can also help the designer
  visualize the accessibility of a destination from different
  locations, and correct any ``blind zone'' with additional signs.
\linespread{1.0}\selectfont
\end{abstract}

}

\maketitle

\IEEEdisplaynontitleabstractindextext

\IEEEraisesectionheading{\section{Introduction}\label{sec:introduction}}
\IEEEPARstart {I}{magine} walking in a subway station with no
wayfinding signs. How could you walk to the right platform after you
buy your ticket? After some random trials, you might finally find
your way to the platform, but this probably would not be a pleasant
experience. You would have saved much time and energy if wayfinding
signs had been placed properly in the environment to guide you
through. A layout with no wayfinding signs is as confusing as a maze.

In ``The VR Book''~\cite{vrbook}, Jerald points out that wayfinding
aids are especially important in virtual environments because it is
very easy to get disoriented throughout a navigation in a virtual
space. A well-constructed environment should include environmental
wayfinding aids thoughtfully put by the level designers, considering
the possible navigation and the navigation goals of the
user. Recently, interesting experiments by Darken and
Peterson~\cite{darken2002spatial} verify that most users would feel
rather uncomfortable being in a largely void virtual environment, and
that it is important to regularly reassure the users that they are not
lost throughout a navigation.

Conventionally, level designers mainly rely on experience or a
``common sense approach''~\cite{commonsense} in creating a wayfinding
design. Given an environment, they think of all likely navigation
scenarios that the user will go through and then place wayfinding
signs or other aids to guide the user accordingly. For example, for a train
station, one common scenario is to walk from the ticket machine,
through the gate, and then to the right platform. Another common
scenario is to walk from the platform to the exit. Directional signs
are then placed along the routes. While this design approach is
straightforward, the efforts required will quickly become daunting
when the number of scenarios scales up as in a real-world
situation. For example, a real-world train station typically involves
tens or more navigation scenarios. Moreover, when placing the signs,
it is necessary to consider the user's visibility and the fact that
the user may miss a sign or make mistakes throughout the
navigation. Designing a wayfinding scheme that jointly considers all
these factors is highly non-trivial and challenging, while a
sub-optimal wayfinding scheme may easily result in a confusing and
frustrating navigation experience of the users.

To tackle these problems, in this work we introduce a novel
computational approach to automatically generate a wayfinding design
for a given environment. To use our approach, the designer simply
specifies all the navigation scenarios likely to be taken by the
users. Our approach will then generate a wayfinding design to
accommodate the needs of all the scenarios while considering a number
of desirable factors relevant to the navigation experience and
management convenience. Through agent-based simulations, our approach
further refines the locations of the wayfinding signs by considering
visibility and robustness with respect to the possible mistakes made
by the users throughout their navigation. After generating a
wayfinding design, the designer can gain further insights of the
design by visualizing the accessibility of a destination from any
other locations in the environment, and remove any blind zones (if
necessary) by adding more signs and re-triggering the optimization.

In a real-world architectural site, typical wayfinding aids include
signs, landmarks and GPS-based mobile navigation system. In a virtual
environment, additional virtual wayfinding aids such as
compasses~\cite{vrbook} and mini-maps~\cite{adams2014fundamentals} can
also be used to facilitate wayfinding. In this work, we focus on
generating signs to guide the user because: 1) signs are a very common
and universal mean for wayfinding; 2) signs as wayfinding aids are
direct yet subtle---the user usually does not need to stop walking
while reassuring his direction with a sign he sees on his way, in
contrast to using other wayfinding aids such as a map which requires
the user to stop his locomotion; 3) signs integrate naturally with
most indoor and outdoor environments.

The major contributions of our work include:

\begin{itemize}
 \item introducing a novel optimization and agent-based approach for automatically generating wayfinding designs.
 \item demonstrating the capability of our approach for generating wayfinding designs for different layouts.
 \item showing how our approach can be further applied for visualizing and editing a generated wayfinding design.
 \item evaluating the effectiveness of our automatically generated
  wayfinding designs in guiding the navigation of users by comparing
  with other wayfinding designs.
\end{itemize}

Additionally, we implement our approach as a handy plugin of the Unity
game engine, which can be used by game level designers to
automatically and quickly generate wayfinding schemes for their
virtual worlds, hence saving their time and manual efforts spent on
determining users' paths and placing wayfinding signs. We will release the
plugin for public use.

\section{Related Work}
\label{sec:related}
To the best of our knowledge, there is no existing work on
automatically generating wayfinding designs for a given layout. We
review some relevant work in wayfinding design for real-world and
virtual environments. We also review some work in sign perception,
navigation and path planning which bring useful insights about the
human factors to consider in a wayfinding design.
\vspace{-3mm}

\subsection{Conventional Wayfinding Design}
We give a succinct overview of the real-world wayfinding design
process, which inspires our computational approach for generating
wayfinding design. 

In architectural design, wayfinding refers to the
user experience of orientation and choosing paths within a built
environment. In the book \emph{The Image of the City}~\cite{lynch},
Lynch defined wayfinding as the ``consistent use and organization of
definite sensory cues from the external environment''. Environmental
psychologists later extended the definition of wayfinding to include
also the use of signage and other graphical and visual clues that aid
orientation and navigation in built environments~\cite{romedi}.

The process of wayfinding involves four major steps~\cite{principles}:
\emph{orientation}, \emph{path decision}, \emph{path monitoring} and
\emph{destination recognition}. \emph{Orientation} refers to
determining one's current location. \emph{Path decision} refers to
selecting paths to navigate to the destination. \emph{Path monitoring}
refers to continuously verifying that the path indeed leads to the
destination. Finally, \emph{destination recognition} refers to
confirming that the destination has been reached. Our goal in this
work is to automatically generate a wayfinding design for a given
environment to facilitate the above wayfinding steps.

\begin{figure}
 \centering
\includegraphics[width=1.0\linewidth]{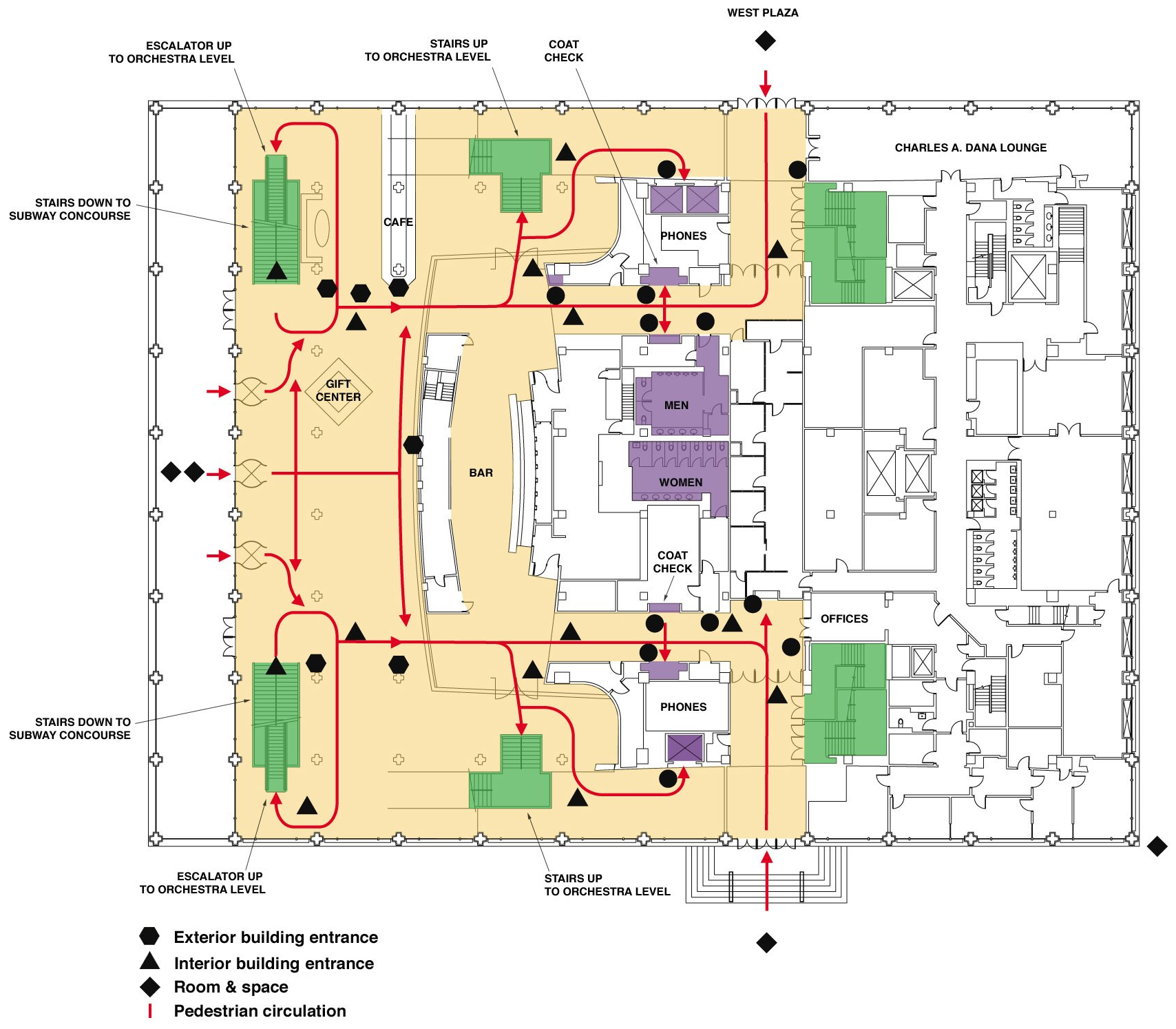}
\caption{{\footnotesize Circulation analysis and wayfinding scheme of
    a concert hall manually created by a wayfinding designers,
    including projected circulation paths and sign types. Courtesy of
    ArchitectureWeek.}}
 \label{fig:circulate}
\end{figure}

Today almost all public spaces and private premises require a
wayfinding scheme~\cite{handbook} to ensure that they are universally
accessible for all users~\cite{act}. To achieve this goal, after a
layout is designed by architects, a wayfinding design team~\cite{swd}
will decide about the wayfinding signs to put in the environment. In
current practice, the design team manually creates a wayfinding scheme
following a ``Common Sense Approach'' mainly based on
experience~\cite{commonsense,common2}. Given a new premise such as a
train station, a wayfinding scheme is designed following these major
steps:

%\begin{enumerate}[leftmargin=0.2cm,itemindent=.5cm]
\begin{enumerate}
\item \emph{Identifying Major Paths:} The design team first identifies
 the major paths likely to be taken by pedestrians, by experience or
 surveys with the property managers. The team examines the site's
 floor plan or make an on-site inspection to estimate people
 flows~\cite{swd}. The goal is to gain a comprehensive understanding
 of the site. \\
 
\item \emph{Devising a Wayfinding Scheme:} Considering all the major
  paths, the design team determines the types and locations of the
  wayfinding signs, which should be placed at an appropriate height
  and angle clearly visible to pedestrians. Additional signs should be
  placed to eliminate any possible confusion caused by
  the architecture itself. As an example, Figure~\ref{fig:circulate} shows the circulation analysis and a wayfinding scheme manually created for a concert hall.\\

\item \emph{Designing, Fabricating and Placing Signs:} After devising
  the sign placement, the designers design the appearance of the signs
  to be manufactured and placed in the
  real environment.\\

\item \emph{Evaluation, Maintenance and Update:} The team maintains
  the wayfinding signs in a database and reviews the sign placement
  periodically to replace any outdated signs.
\end{enumerate}

Interested readers may refer to the
literature~\cite{swd,signsystem,handbook} for more detail of the
design process. Similar to the real-world wayfinding design process,
our computational approach focuses on automatically identifying
locations for placing signs in an environment according to the
designer-specified navigation goals of the pedestrians.

\subsection{Wayfinding Design for Virtual Environments} 

% 1) wayfinding design is important consideration for virtual environments
% 2) what kind of wayfinding cues are used? how level designer usually generates them?
% 3) destination maps. Maps + wayfinding signs comparison. Good to have both.
% 4) affordance aware procedural modeling of layout design. e.g., furniture, lights, crowd driven layout. We are procedural modeling too, placing signs.

\begin{figure}
\centering
\subfloat[Fallout 4]{\includegraphics[width=0.48\linewidth]{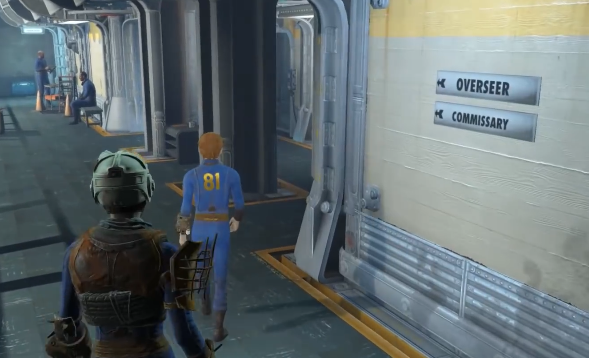}}
\hspace{1mm}
\subfloat[The Elder Scrolls IV: Oblivion]{\includegraphics[width=0.466\linewidth]{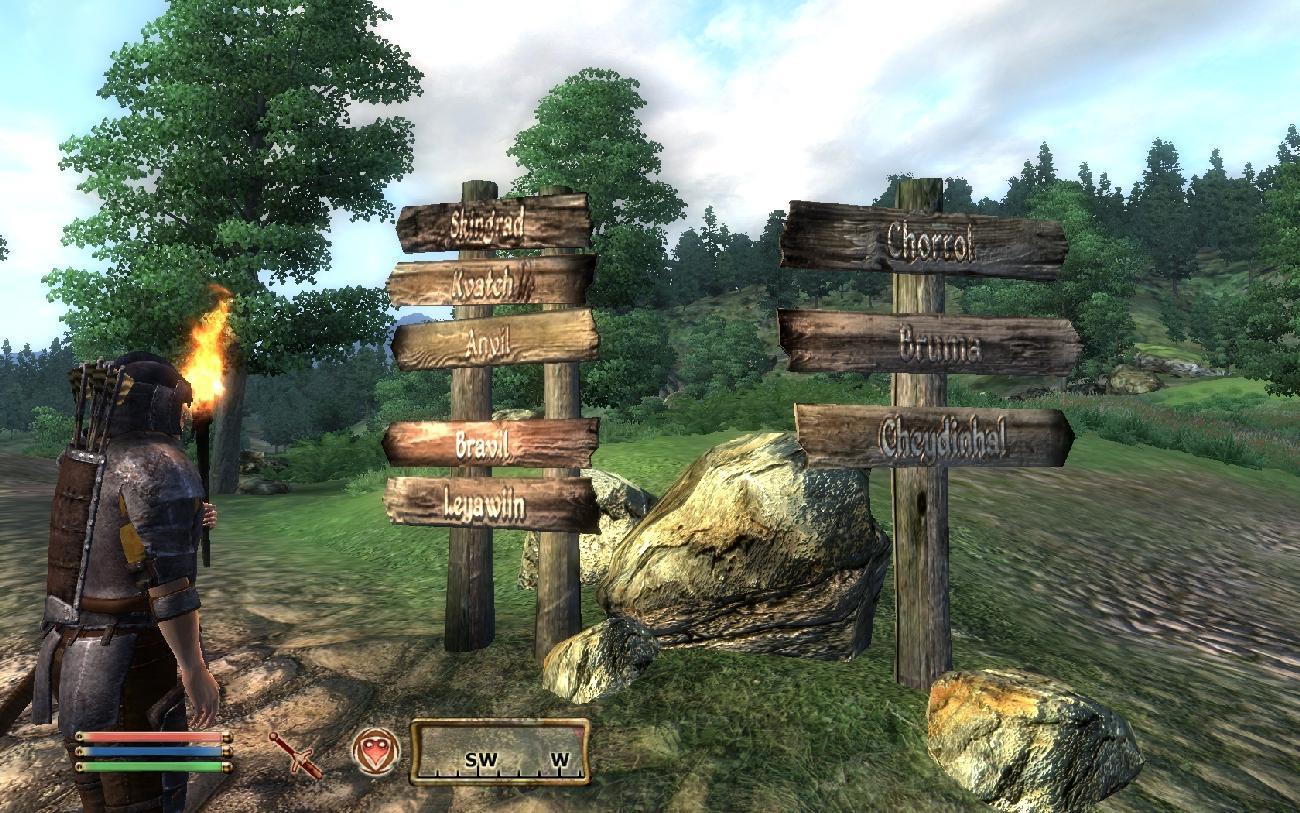}}
\caption{{\footnotesize Example road signs used in video games.}}
 \label{fig:gameroadsign}
\vspace{-4mm}
\end{figure}

Wayfinding aids are crucial in virtual environments because they help
users form cognitive maps, maintain a sense of position and direction
of travel, and find their ways to their
destinations~\cite{vrbook}. Common wayfinding aids in virtual
environments include signs, maps, landmarks, light, and
paths~\cite{elvins1997worldlets,vrbook}. In designing a highly
immersive and steerable virtual environment, it is important for level
designers to make use of wayfinding aids effectively to enhance
spatial understanding of the environment so that users can comprehend
and operate smoothly~\cite{mitchell2012game,vrbook}. This principle
also applies to game level design. In his book, game designer Michael
Salmond emphasizes the use of a road sign system in games as an
important wayfinding tool to provide players with a highly immersive
navigation experience~\cite{videogame}. \Fref{fig:gameroadsign} shows
some example road signs used in the popular video games Fallout 4 and
the Elder Scrolls IV: Oblivion.

In current practice, wayfinding aids are manually added to a virtual
environment by level designers and then empirically tested for
effectiveness, which depends on the quantity and quality of wayfinding
aids provided to users, yet research found that it could be
overwhelming to users if exposed to too many wayfinding
aids~\cite{bowman20043d}.

Darken and Sibert conducted an important
study~\cite{darken1996wayfinding} about the wayfinding strategies and
behaviors of human users in large virtual worlds. Their experiments
verified that human wayfinding strategies and behaviors in large
virtual worlds are strongly influenced by environmental cues. Their
experiments asserted that \emph{humans generally adopt physical world
wayfinding strategies in large virtual worlds, hence common wayfinding
aids in the physical world can be effectively applied to facilitate
wayfinding in virtual worlds}. Based on the insights, Cliburn and
Rilea~\cite{cliburn2008showing} conducted a further study to compare
human performance in searching for an object in a virtual environment
with no aid present, with maps and with signs. The results show that
subjects who navigated the virtual environment with the aid of signs
achieve superior performance than under other conditions. These findings
motivate us to investigate the automatic propagation of directional
signs in virtual environments to enhance wayfinding.

\ssection{Wayfinding Map Generation.} In computer graphics, there are
interesting approaches for automatically generating tourist
brochures~\cite{birsak2014automatic} and destination
maps~\cite{kopf}. Though these maps are intended for real-world
navigation use, they could potentially be used to assist navigation in
virtual environments. Given a map and some desired destinations, these
approaches select a subset of roads to reach the destinations, and
visualize the important routing instructions on a generated map which
is intuitive to use. Our approach is inspired by these approaches, but
focuses on optimizing the placement of wayfinding signs in the layout
so as to guide pedestrians to reach their destinations
easily. Combining automatically generated maps with the wayfinding
signs generated by our approach can potentially provide users with
effective wayfinding aids to navigate smoothly in virtual
environments.

% Quincy comments
% \blue{
% \noindent [***\\
% I am wondering that if some reviewers would question that why we do not use the shortest path for navigating in both real and virtual environments. How do we defend this point?

% Can we say:
%   Because most people, I believe, would prefer to along the shortest local path, if they know the shortest local path, rather than along the farther one (optimized) to reach the destination. Indeed, the optimal signs placement will make the local path little longer, but it also reduces the number of the road signs which means reducing the building and maintaining fee of the road signs placement. Additionally, the layout designers can focus on developing the critical buildings along the paths of the optimal road signs placement, so that improve the use rate of roads and save the time on re-designing the road signs. (Let's say, for a given wayfinding scheme about a touring city, there are scenic spots, a bus station, and a hotel. Then, we are going to build a restaurant to satisfy tourists need; we hope these relevant POIs stay closely and convenient for the tourists. One of the best ways is to put the restaurant along the optimal paths between the existed POIs.)  On the other hand, in the MMO games (Massively Multiplayer Online games), the level designers also can focus on placing the monsters or the task locations along the path of the road signs placement so that to guide the players to gather together and encourage more interactive activities between the players.
% \\
% \noindent ***]
% } % end blue

\subsection{Perception, Path Planning and Navigation}
Our wayfinding design approach is also inspired by how humans perceive and
navigate in everyday environments.

\ssection{Perception.} In everyday environments, humans continually
shift their gaze to retrieve wayfinding cues for making navigation
decisions~\cite{H1,H2,H3}. Human visual attention is known to be
attracted by low-level features such as changes in color, intensity,
orientation and contrast~\cite{itti}, and by high-level scene
context~\cite{H3}. Some particular categories of objects, such as
signs and texts~\cite{H4,H5,A1}, are known to strongly attract eye
fixations regardless of their low-level visual saliency. Therefore, we
focus on optimizing the placement of wayfinding signs in our approach.

% \blue{
% \noindent In our approach, we focus on optimizing the road signs placement
% %of arrow and textual signs 
% to guide pedestrians to their destinations 
% %effectively.
% accordingly.
% }

\begin{figure*}[t]
 \centering
%\subfloat[Initialization]
{\includegraphics[width=0.99\linewidth]{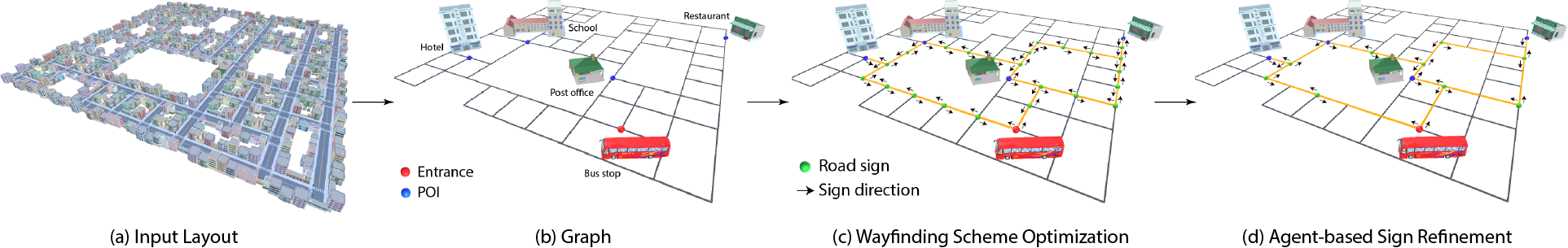}}
\caption{Overview of our approach. (a--b) An input layout is first converted into a graph representation. (c) A wayfinding scheme optimization is then performed to find the optimized paths from the entrance to the points of interests, considering wayfinding criteria such as path lengths, angles and number of decision points. (d) An agent-based sign refinement subsequently takes place, which employs agent-based simulations to evaluate and refine the sign placement according to the navigation properties of agents mimicking human pedestrians.}
 \label{fig:workflow}
\vspace{-4mm}
\end{figure*}

\ssection{Path Planning and Navigation.} Given a layout, there are
usually multiple paths a pedestrian can take to navigate from a
starting point to a destination. For instance, suppose a hiker wants
to walk from the bottom to the top of a hill. He may walk a path which
is mostly straight, or a shortcut with sharp turns. A common strategy
for path planning is to design a cost function to evaluate each path,
and then search for a path that corresponds to a low
cost~\cite{jaillet2010sampling,nilsson2014principles,galin2010procedural,duckham2003simplest}. For
a low-dimensional configuration space, a grid-based search such as
A*~\cite{hart1968formal} or D*~\cite{stentz1994optimal} can be applied
to find an optimal path. For a high-dimensional configuration space,
sampling-based
approaches~\cite{kavraki1996probabilistic,miao2013dynamic} are
commonly applied to find an optimal or near-optimal solution.

For path planning, common navigation factors to consider in the cost
function involve: 1) path length (one wants to choose a short path to
reduce the travel time needed to reach the destination); 2) number of
turns (one wants to minimize the number of turns to reduce the
complexity of the route~\cite{winter2002modeling,kopf}); 3) number of
decision points (each intersection is a decision point where the
pedestrian will need to decide which road to follow next; one wants to
minimize the number of decision points to reduce the chance of making
mistakes). Arthur and Passini~\cite{passini1992wayfinding} noted that
the number of decision points has an important influence on the
difficulty of performing wayfinding. Casakin et al. conducted
empirical studies~\cite{casakin2000schematic} which further verify
these observations. We consider these criteria in the generation of
our wayfinding designs, after which we will place the wayfinding signs
and refine the placement based on agent's properties. Moreover, the
designer can control the importance of each criterion by adjusting its
associated weight. Our approach will generate a wayfinding design
accordingly.

\ssection{Navigation Mistakes.} Humans occasionally make mistakes in
navigation. For example, it is common for pedestrians to miss a sign
due to occlusion by other pedestrians; distractions such as
advertisements or events happening in the
environment~\cite{nassar2008assessing,burns1998wayfinding}; or a
wrongly recognized sign or landmark~\cite{golledge1999human}. It is
also common for pedestrians to make wrong turns in
navigation~\cite{burns1998wayfinding}. A well-thought-through
wayfinding design should tolerate these kinds of human
mistakes~\cite{darken1996wayfinding,passini1992wayfinding}---a
pedestrian should still be able to reach to his destination even if
he makes mistakes occasionally.

\ssection{Agent-based Evaluation.} Martin
Raubal~\cite{raubal2001human} used agent-based simulation to evaluate
human wayfinding in unfamiliar environments, yet the simulation used
does not consider the mistakes that can be made by the agents;
further, it is unsure how such evaluations can be used to enable
automatic sign placement. In contrast, our agent-based simulations
consider navigation mistakes and we also show how such simulations can
be used to create a robust wayfinding design. Our approach is also
motivated by autonomous agents~\cite{shao2005autonomous} and crowd
simulations~\cite{treuille2006continuum,pelechano2008virtual,thalmann2007crowd}.
However, instead of generating realistic agent simulations, we focus
on applying agent-based simulations for optimizing wayfinding designs.

To achieve a robust wayfinding design, our approach conducts
agent-based simulations in placing the signs to evaluate how well the
design can tolerate occasional mistakes made by agents. Using our
approach, the designer can control how robust the generated wayfinding
design needs to be by changing the agent parameters. For example, in
creating the wayfinding design for a subway station where the
pedestrians (many of whom are first-time visitors) are generally
expected to be unfamiliar with the environment, the designer can
adjust the agents to have a higher chance in making mistakes. Our
approach will generate a more robust wayfinding design by placing
signs in important locations so that pedestrians can still find their
ways despite the mistakes.

\subsection{Computational Layout Design}
Layout design is an important problem in computer graphics. A layout
typically consists of a number of sites connected by paths, with each
site serving a different purpose. Computer-generated layouts can be
used for creating virtual environments where virtual agents and human
users can navigate for simulation and entertainment purposes. Galin et
al. proposed to generate roads procedurally given a natural landscape
with river and hills~\cite{galin2010procedural}. Computationally
generated layouts can also be used for architectural
design~\cite{merrell2010computer,bao2013generating,peng2014computing}
and urban
planning~\cite{vanegas2012procedural,vanegas2009interactive,vanegas2012inverse,func}. Refer
to the survey~\cite{smelik2014survey} by Smelik et al. for a
comprehensive review of the state-of-the-art procedural modeling
techniques for generating layout designs for virtual environments.

An important consideration in designing a layout is the navigation
experience of the pedestrians. Recently, Feng et al. proposed an
approach~\cite{midlayout} which uses crowd simulation to generate
mid-scale layouts optimal with respect to human navigation properties
such as mobility, accessibility and coziness. However, concerning
navigation, one important consideration is missing: the wayfinding
experience of the pedestrians in the generated environments. We argue
that their generated layouts are navigation-aware only if wayfinding
signs are properly placed in the layouts.

% An important consideration in designing a layout is the navigation
% experience of the pedestrians. A layout should be designed in such a
% way that the pedestrians can conveniently walk to their desired sites
% without experiencing any discomfort due to congestion. To achieve this
% goal, recently Feng et al. proposed an approach~\cite{midlayout} which
% uses crowd simulation to generate mid-scale layouts which are optimal
% with respect to human navigation properties such as mobility,
% accessibility and coziness. They demonstrated their approach in
% generating crowd-aware layouts for shopping malls, theme parks and
% train stations, etc. However, concerning navigation, one important
% consideration is missing in their approach: the wayfinding experience
% of the pedestrians in the generated environments. In their crowd
% simulation, they assume that the agents representing pedestrians can
% perfectly navigate to their destinations in a given layout without the
% help of any wayfinding aid, while in reality this is generally not the
% case. We argue that their generated layouts are navigation-aware only
% if proper wayfinding signs are also placed in the generated layouts.

In this regard, we consider our automatic wayfinding design
approach as complementary to automatic layout design or road network
generation approaches. The wayfinding signs automatically generated by
our approach can enhance the navigation experience of users in virtual
environments, as we show in our experiments.

\section{Overview}
\label{sec:overview}
Figure~\ref{fig:workflow} shows an overview of our approach. We use a
layout called \emph{City} as our illustrative example to describe our
approach. Our approach works on a graph representing an input
layout. It consists of two major steps: \emph{Wayfinding Scheme
  Optimization} and \emph{Agent-based Sign Refinement}. In the
\emph{Wayfinding Scheme Optimization} step, our approach determines
the paths for pedestrians to walk from the starting points to the
destinations under different navigation scenarios specified by the
user. Different human-centered navigation criteria such as turning
angles and walking distances are jointly considered through an
optimization to determine the paths to take. In the \emph{Agent-based
  Sign Refinement} step, our approach places wayfinding signs
strategically at appropriate locations along the paths. By using
agent-based simulations to evaluate sign placement, our approach takes
into account different human properties such as visibility and the
possibility of making navigation mistakes. Depending on the
requirements of the navigation scenarios, the designer can easily generate
a wayfinding design that satisfies the domain-specific requirements,
by changing the weights of different criteria in the wayfinding scheme
optimization and the parameters of the agent-based simulations.

%\begin{figure}
% \centering
%\includegraphics[width=0.9\linewidth]{newversion/wayfindingOPT}
%\caption{{\footnotesize Wayfinding Scheme Optimization.}}
% \label{fig:wayfindingOPT}
%\end{figure}

%\begin{figure}
% \centering
%\includegraphics[width=0.9\linewidth]{newversion/agenbasedRSP}
%\caption{{\footnotesize Agent-based Sign Placement.}}
% \label{fig:agenbasedRSP}
%\end{figure}

% [Story of running example]
%\blue{In a given city layout, there are too many POIs need to be guided to 
%and limited space for placing road signs at each intersection(cite XXX paper?). In order to 
%full use of limited space, every wayfinding scheme should has a meaningful 
%setting. In our running example, it shows a scheme setting, for academic visiting.
%There are the bus stop, an entrance for entering the city; 
%the university, the important place for this scheme setting; 
%a post office, a place which can mail a postcard or souvenirs; 
%the most famous local restaurant, most visitors want to try; 
%a hotel, a living place which closest to the university. 
%These five places consist of a wayfinding scheme for most academic visiting use.
 %}

\section{Problem Formulation}
\label{sec:formulation}

\subsection{Representation}
\label{sec:representation}

\ssection{Graph Construction.} To apply our approach, the user first
%manually 
creates a graph $G=\{V, E\}$ to represent the input layout, where $V$
is the set of nodes representing the intersections, entrances and
points-of-interest (POIs), and $E$ is the set of edges representing
the connecting paths between adjacent nodes. The creation process is
simple and is similar to specifying a waypoint system in typical game
level design. The user places nodes at the intersections, entrances
and POIs of the layout. For example, in the illustrative example,
\emph{City}, the POIs include the school, the post office and so
forth. The user also adds an edge between two adjacent nodes if the
places represented by the nodes are connected by a road.

\ssection{Source-Destination Pairs.} A source-destination pair encodes
a navigation scenario to be considered by our approach, \eg, going
from a bus stop to a restaurant, akin to an input pair a wayfinding
designer creates to specify a navigation scenario in conventional
wayfinding design~\cite{swd}. Each pair $z_i = (s_i,d_i)$ consists of
a source (starting point) $s_i$ and a destination $d_i$.

To facilitate the creation of source-destination pairs, by default our
approach automatically generates a source-destination pair between
every node representing an entrance and every node representing a POI,
with the former being the source and the latter being the
destination. Additionally, the user can specify any extra pair if
needed. For instance, in the \emph{City} example, he may want to create
a pair connecting the hotel and the restaurant.

\ssection{Importance Values.} We also allow the user to assign an
importance value $\kappa_i \in [0,1]$ to each source-destination
pair. For instance, in the \emph{City} example, the (\emph{Hotel},
\emph{Restaurant}) pair can be given a higher importance value if many
pedestrians are expected to walk from the \emph{Hotel} to the
\emph{Restaurant}, whereas the (\emph{School}, \emph{Restaurant})
pair can be given a lower importance value if fewer pedestrians are
expected to walk from the \emph{School} to the
\emph{Restaurant}. In the optimization, the path connecting the
\emph{Hotel} with the \emph{Restaurant} should be given a higher
priority, compared to the path connecting the \emph{School} to
the \emph{Restaurant}. If a trade-off exists, it is important to make
sure that pedestrians can walk conveniently from the \emph{Hotel} to
the \emph{Restaurant}, while it may not matter as much for pedestrians
to walk a somewhat inconvenient path from the \emph{School} to
the \emph{Restaurant}.

\section{Wayfinding Scheme Optimization}
\label{sec:optimization}

Given a source-destination pair $z_i=(s_i, d_i)$, there could exist multiple
possible paths from $s_i$ to $d_i$. Let $P_{z_i}$ denote the
set of all such paths. Our goal in this step is to generate a
wayfinding scheme that takes all source-destination pairs $\{z_i\}$
into account and selects a path for each pair. In other words, we
select a path $p_i \in {P}_{z_i}$ for each pair $z_i$, such that the
set of all selected paths $P=\{ p_i \}$ satisfies some local and
global criteria defined by our cost terms. We formulate our problem as
an optimization of a total cost function:

\begin{eqnarray}
C^{\textrm{P}}_{\textrm{all}}(P) = w^{\textrm{L}}_{\textrm{local}} C^{\textrm{L}}_{\textrm{local}}+
w^{\textrm{N}}_{\textrm{local}} C^{\textrm{N}}_{\textrm{local}}+
w^{\textrm{A}}_{\textrm{local}} C^{\textrm{A}}_{\textrm{local}}+ \\ \nonumber 
w^{\textrm{L}}_{\textrm{global}} C^{\textrm{L}}_{\textrm{global}}+
w^{\textrm{N}}_{\textrm{global}} C^{\textrm{N}}_{\textrm{global}}
\label{eqn:total_cost}
\end{eqnarray}

The total cost function $C^{\textrm{P}}_{\textrm{all}}(P)$ refers to a weighted sum of cost terms
encoding the length, number of decision points and the amount of turns
of each path, as well as the length and number of decision points of
the overall wayfinding scheme. The user can adjust the importance of
different design criteria by changing the weights of the
corresponding cost terms, to accommodate the domain-specific needs of the layout for
which the wayfinding scheme is designed. We describe each cost term in
detail as follows.

\subsection{Wayfinding Cost Terms}
\label{sec:costs}

\ssection{Local Path Length.} In general, pedestrians prefer to walk
a short distance~\cite{darken1996wayfinding,swd,foltz1998designing}. Hence, for each source-destination pair, a shorter
path is preferred. We define a cost
to penalize the length of the selected path of each source-destination
pair:

\begin{equation}
C^{\textrm{L}}_{\textrm{local}}(P) = \frac{1}{|P| L_E} \sum_{p \in P} \kappa_p L(p),
\end{equation}

\noindent where $|P| L_E$ is the normalization factor with $|P|$
being the number of source-destination pairs and $L_E$ being the
total length of all edges in graph $G$. $L(p)$ returns the length of
path $p$. $\kappa_p \in [0,1]$ is the importance value assigned to the
source-destination pair that path $p$ belongs to.

\ssection{Local Path Node.} The nodes in our formulation correspond to
\emph{decision points} in the wayfinding
literature~\cite{swd}. Decision points are locations where pedestrians
need to make a decision about which direction to go, such as an
intersection between paths (\eg, a lobby in a subway station); or
where pedestrians need to confirm the identity of the current
location, such as a place of interest (\eg, a platform in a subway
station). Directional or identification signs need to be placed at
decision points to guide pedestrians to find their
directions~\cite{swd,signsystem} or identify their current
locations. Paths with lots of decision points should be
avoided~\cite{passini1992wayfinding} as making each navigation
decision induces stresses to the pedestrians for the fear of making a
wrong decision that may lead to a wrong
place~\cite{passini1992wayfinding,payne2009understanding}. Therefore
we define a cost to penalize the number of decision points of each
path:

\begin{equation}
C^{\textrm{N}}_{\textrm{local}}(P) = \frac{1}{|P| |V|} \sum_{p \in P} \kappa_p N(p),
\end{equation}

\noindent where $|P| |V|$ is the normalization factor with $|P|$
being the number of source-destination pairs and $|V|$ being the total
number of nodes in graph $G$. $N(p)$ returns the total number of nodes
along path $p$.

\ssection{Local Path Angle.} Research in spatial
orientation~\cite{golledge1999human} suggests that paths with varying
orientation tend to confuse pedestrians in wayfinding, causing
disorientation, anxiety and discomfort~\cite{darken2002spatial}. A
wayfinding scheme composed of straight paths is more intuitive for
navigation~\cite{duckham2003simplest}. We therefore include a cost
term to penalize the selection of paths with varying orientation:

\begin{equation}
C^{\textrm{A}}_{\textrm{local}}(P) = \frac{1}{|P| |V| \pi} \sum_{p \in P} \kappa_p A(p),
\end{equation}

\noindent where $|P| |V| \pi$ is the normalization factor with $|P|$
being the number of source-destination pairs and $|V|$ being the total
number of nodes in graph $G$. The maximum absolute turning angle
between two adjacent edges is $\pi$. $A(p)$ returns the sum of
absolute turning angles between all adjacent edges along path $p$.

% \begin{equation}
% C^{\textrm{A}}_{\textrm{local}}(P) = \frac{1}{|P| \pi} \sum_{p \in P} \kappa_p \frac{A(p)}{N(p)-2},
% \end{equation}

% \noindent where $|P| \pi$ is the normalization factor with $|P|$ being
% the number of source-destination pairs. The maximum turning angle
% between two adjacent paths is $\pi$. $A(p)$ returns the sum of
% absolute turning angles between all adjacent paths along path
% $p$. $N(p)$ returns the total number of nodes along path $p$. Note
% that $A(p)$ is only defined if $N(p)>2$. For a path with only $2$
% nodes (\ie, a source and a destination), our approach just sets its
% angle cost term to be zero.

\ssection{Global Path Length.} Our approach encourages paths to
overlap with each other so as to minimize the total length of roads
(edges) that are part of a path. This property could be useful from the
management's perspective~\cite{swd,romedi}, because by directing the
flow of human movement to fewer roads, fewer roads will need to be
maintained, patrolled and lightened up. We define a cost to encourage
overlapping paths accordingly:

\begin{equation}
C^{\textrm{L}}_{\textrm{global}}(P) = \frac{L(P)}{L_E},
\end{equation}

\noindent where $L_E$ is the total length of all edges in graph $G$ as the
normalization factor. $L(P)$ returns the total length of the edges that
belong to any path in $P$.

\ssection{Global Path Node.} Our approach also encourages different
paths to share nodes. Similar designs can be observed in the
wayfinding schemes of different real-world premises, such as subway
stations, shopping malls and concert halls, where people are directed
to a lobby or an information desk that can lead to multiple
destinations (see Figure~\ref{fig:circulate} for an example). From the
management's perspective, it could be easier to maintain signs
centralized at certain locations in the
environment~\cite{handbook,signsystem}. Also, centralizing signs could
save space, which could be reserved for other better
uses~\cite{foltz1998designing}. We define a cost to encourage node
sharing accordingly:

\begin{equation}
C^{\textrm{N}}_{\textrm{global}}(P) = \frac{N(P)}{|V|},
\end{equation}

\noindent where $|V|$ is the total number of nodes in graph $G$ as the
normalization factor. $N(P)$ returns the total number of nodes that
belong to any path in $P$.

\begin{figure*}
\centering
\subfloat[Initialization]{\includegraphics[width=0.25\linewidth]{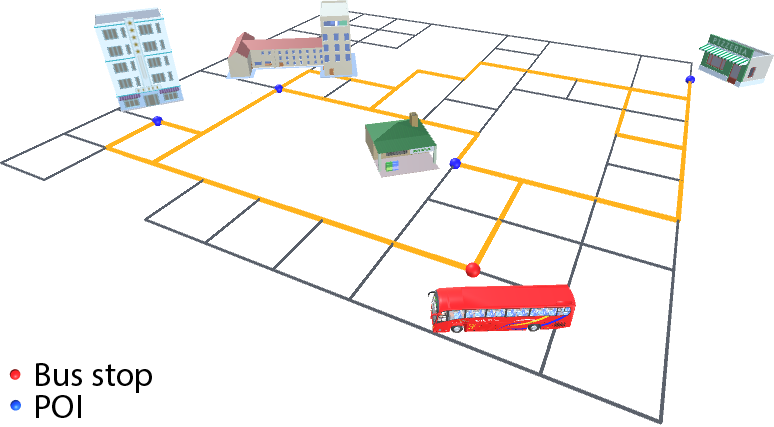}}
%\subfloat[Iteration 1,000]{\includegraphics[width=0.2\linewidth]{newversion/ai_iterative_1000}}
\subfloat[Iteration 5,000]{\includegraphics[width=0.25\linewidth]{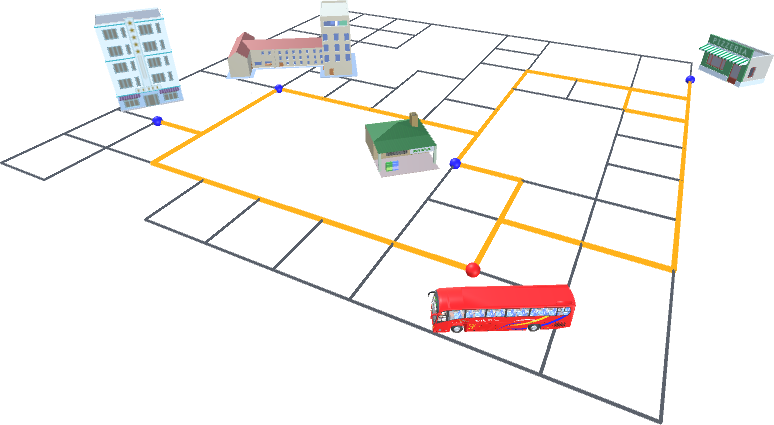}}
\subfloat[Iteration 30,000]{\includegraphics[width=0.25\linewidth]{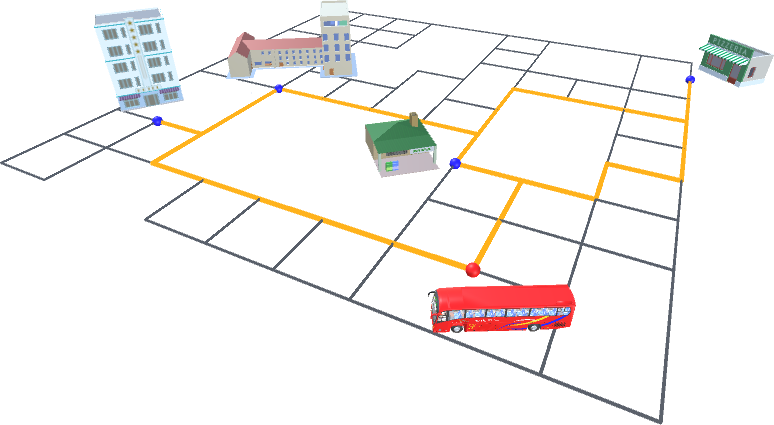}}
\subfloat[Iteration 40,000 (result)]{\includegraphics[width=0.25\linewidth]{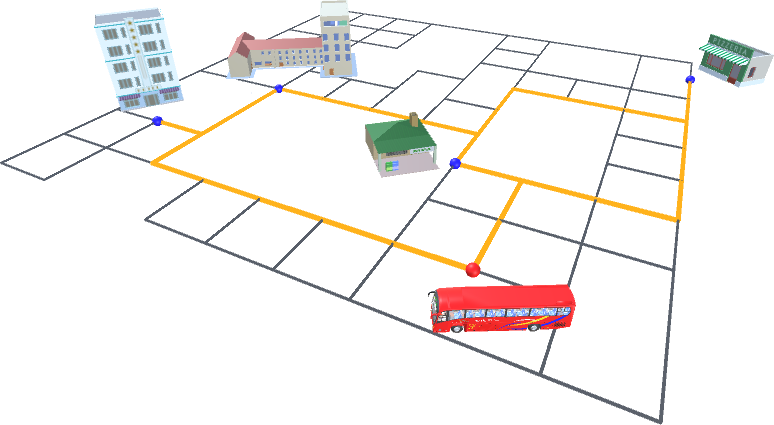}}
%\subfloat[Running cost]{\includegraphics[width=0.2\linewidth]{newversion/running_cost_fig}}
\\
\caption{{\footnotesize Wayfinding schemes generated over the
    iterations of an optimization of the illustrative example,
    \emph{City}. (a) Initialization. The source-destination pairs
    include walking from the bus stop to each POI, and walking between
    every pair of POIs. The path of each pair is randomly chosen from
    its $k$ shortest paths. (b) Iteration 5,000. The selected paths
    start to overlap. (c) Iteration 30,000. The result is still
    sub-optimal. For example, the path connecting the bus stop to the
    restaurant at the upper right still shows large turning angles and
    consists of an excessive number of nodes (decision points). (e)
    Optimized result.}}
 \label{fig:iterations}
\vspace{-3mm}
\end{figure*}

\subsection{Optimization}
\label{sec:opt}
For each source-destination pair $z_i$, there exist a lot of possible
paths going from the source to the destination. For instance, pair
(\emph{Bus Stop}, \emph{School}) in the illustrative example
(Figure~\ref{fig:workflow}) has more than $1,000$ possible
paths. Given the many combinations of possible paths of all pairs, the
solution space could be huge as it grows exponentially with the number
of pairs being considered.

To reduce the search space for a solution, we devise a sampling-based,
stochastic search algorithm to solve the optimization problem as
follows. For each pair, we only consider the first loopless $k$
shortest paths, which can be found by Yen's
algorithm~\cite{yen1971finding} in $O(|E|+|V|\log(|V|))$ time using a
Fibonacci heap, where $|E|$ is the number of edges and $|V|$ is the
number of nodes. Deviation algorithms~\cite{de1999deviation} and
alternative implementations \cite{martins2003new} exist that could
further enhance computational efficiency, yet we adopt the
classical implementation for simplicity.

Given the $k$ shortest paths for each source-destination pair, we find
a combination of paths of all source-destination pairs which
corresponds to a low cost value. Even though we reduce the size of the
solution space this way, an exhaustive search for the global optimum
would still require heavy computation exponential to the number of
pairs being considered.

\begin{figure}
 \centering
\includegraphics[width=0.9\linewidth]{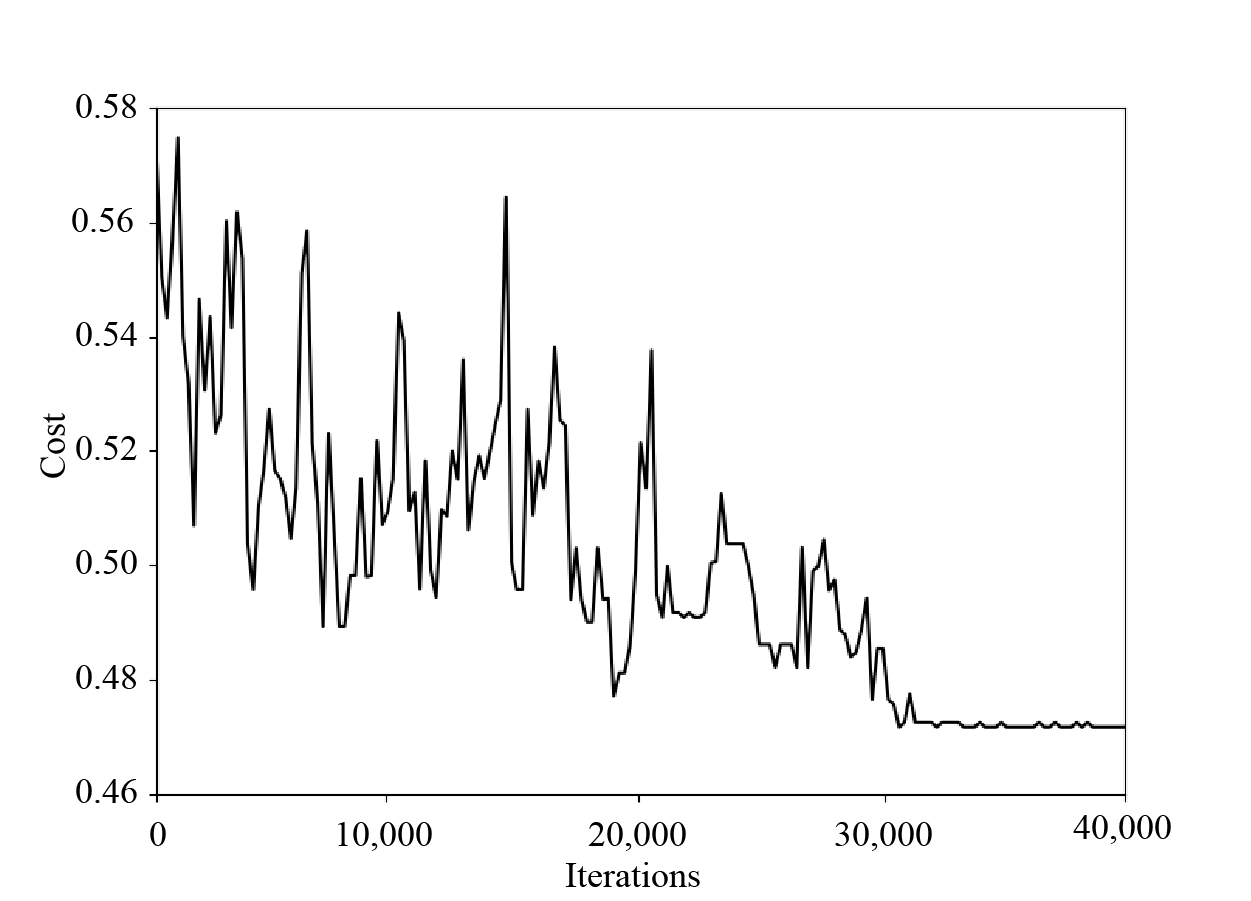}
\caption{{\footnotesize Cost throughout the optimization of the \emph{City} example.}}
 \label{fig:runningcost}
\vspace{-7mm}
\end{figure}

Instead, our approach finds a local optimum as an approximate
solution. We apply the simulated annealing
technique~\cite{Kirkpatrick83optimizationby} with a Metropolis
Hasting~\cite{Metropolis1953} state-searching step to explore the
complex optimization landscape. The optimization proceeds
iteratively. In each iteration, the current solution $P$ is altered by
a proposed move to another solution $P'$, which may or may not be
accepted depending on the acceptance probability of the proposed
solution. More specifically, the acceptance probability is calculated
by the Metropolis criterion:

\begin{equation}
\Pr(P' | P) = \min (1, e^{\frac{1}{T} (C^{\textrm{P}}_{\textrm{all}}(P) - C^{\textrm{P}}_{\textrm{all}}(P')) } ),
\label{eqn:accept_prob}
\end{equation}

\noindent where $T$ is the temperature of the annealing process. $T$
is high at the beginning of the optimization, allowing the optimizer
to explore the solution space more aggressively; $T$ is low towards
the end of the optimization, allowing the optimizer to refine the
solution. Essentially, the optimizer accepts any solution with a lower
cost, while it accepts a solution with a higher cost at a probability:
the higher the cost, the lower the acceptance probability. The
optimization terminates if the absolute change in cost is less than
$1\%$ over $1,000$ iterations.

\begin{figure}
\centering
\subfloat[]{\includegraphics[width=0.48\linewidth]{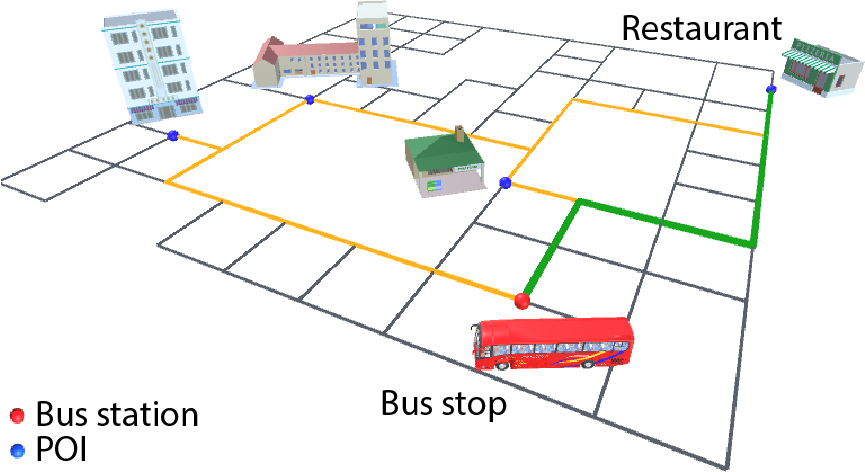}}
\hspace{1mm}
\subfloat[]{\includegraphics[width=0.48\linewidth]{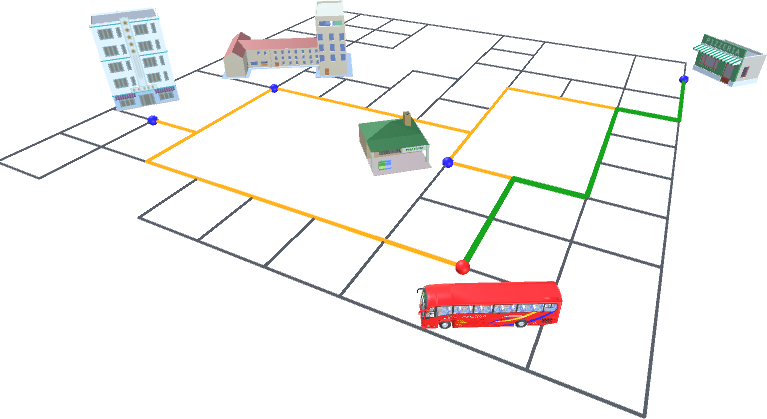}}
\caption{{\footnotesize Experimenting with different importance values. (a) Wayfinding scheme generated with all source-destination pairs having the same importance value of $0.5$. The path from the \emph{Bus Stop} to the \emph{Restaurant} is shown in green. (b) Wayfinding scheme generated with the (\emph{Bus Stop}, \emph{Restaurant}) pair having a lower importance value of $0.25$. In this case, the global path length is improved (\ie, the global path length is shorter) while the local path angle of the path from the \emph{Bus Stop} to the \emph{Restaurant} becomes worse (\ie, the path has more orientation changes).}}
\label{fig:importance}
\vspace{0mm}
\end{figure}

Figure~\ref{fig:iterations} shows the wayfinding schemes generated
over the iterations of the optimization process of the illustrative
example. Figure~\ref{fig:runningcost} shows the decay in cost over the
optimization process. We also experimented with changing the
importance values of the source-destination pairs; the resulting
wayfinding schemes are depicted in Figure~\ref{fig:importance}.

% \ssection{Formulation.} The overall formula.
% \ssection{Graph Generation.} How to convert from layout to graph.
% \ssection{Hypothesis Generation.} Talk about k-shortest paths.
% \ssection{Paths Evaluation.} Talk about the combinatorial optimization. If many combinations, then do mcmc to optimize faster.
% \ssection{Map Placement.} Talk about the optimization.

\ssection{Proposed Moves.} Our proposed moves follow a simple
design. Depending on the number of source-destination pairs $|P|$, our
optimizer changes the selected paths of up to $|P|$ source-destination
pairs in a single move. The probability $\Pr_x$ of drawing a move to
change the selected paths of $x$ pairs is inversely proportional to
$x$, \ie, $\Pr_x = \frac{|P|-x+1}{X}$, where $X=\sum^{|P|}_{i=1}i$. A
selected path is randomly changed to another path from the set of $k$
shortest paths of its corresponding source-destination pair.

\ssection{Parameter Settings.} In our experiments, we initialize $P$
by randomly selecting a path from one of the $k$ shortest paths for
each source-destination pair. By default we adaptively set $k$ such
that the length of the $k$-th shortest path is just within $16\%$ of
the length of the first shortest path, as research in spatial
cognition finds that humans typically choose a path with a length
within $16\%$ of that of the shortest
path~\cite{duckham2003simplest}. Unless otherwise specified, each pair
is assigned the same importance value $\kappa_i=\frac{1}{|P|}$, and we
empirically set the weights $w^{\textrm{L}}_{\textrm{local}}$ and
$w^{\textrm{N}}_{\textrm{local}}$ to $1$,
$w^{\textrm{L}}_{\textrm{global}}$ and
$w^{\textrm{N}}_{\textrm{global}}$ to $5$, and
$w^{\textrm{A}}_{\textrm{local}}$ to $10$. These parameters and weights
can be adjusted via the interface of our tool according to
domain-specific design needs---a flexibility provided by our
optimization-based design framework.

% The above parameters can be adjusted according to the requirements in
% designing a wayfinding scheme for a specific scenario --- a flexibility
% provided by our approach. 
%The optimization terminates when the decrease
%in total cost value is less than $3\%$ over the past $50$ iterations.

% Figure~\ref{fig:iterations} shows the iterations over the optimization
% of the illustration example. Figure~\ref{fig:only} shows the optimized
% solutions obtained by considering each cost term. Note that each
% optimized solution is different. Optimizing the total cost function
% is equivalent to reaching a compromise over all weighted cost terms.

% (The cost function contains a large number of variables (the node
% locations) and includes both nonlinearities and discontinuities (e.g.,
% e int ). Thus, computing derivatives for optimization via gradient
% descent is difficult.)

% Overview (no title)

% Optimization
% - Overview, math representation (no title)
% - Cost Terms

% - Agent-based Evaluation
%  - Agent Model
%  - Simulation

% -  Iterative Refinement
%  - (initialization)
%  - general process
%  - moves
%  - termination conditions

\section{Agent-based Sign Refinement} 
\label{sec:agent}

The wayfinding scheme optimization in the previous step produces a
wayfinding scheme which comprises paths from the sources to the
destinations. In this section, we discuss how our approach
automatically places signs for each path to facilitate wayfinding.

\subsection{Overview}
\setlength{\columnsep}{5pt}
\begin{wrapfigure}[14]{r}{0.22\linewidth}
\vspace{-3mm}
\includegraphics[width=1.0\linewidth]{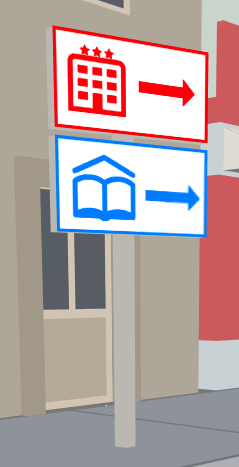}
\caption{{\footnotesize Example sign.}}
\label{fig:sign3d}
\end{wrapfigure}Each node along a path corresponds to a decision point where a sign
may be placed~\cite{swd}. In our experiment design, a sign shows an arrow
pointing to the next node and the destination's name or symbol. Two or
more signs placed at the same node are combined into a single sign
showing multiple pieces of wayfinding
information. Figure~\ref{fig:sign3d} shows an example sign placed at a
street corner in the \emph{City} scene.

A trivial yet unrealistic solution is to place a sign showing the
direction to the destination at every node along a path, so that a
pedestrian walking along the path will keep reassured that he is
heading to his destination. However, this solution will involve
placing many redundant signs occupying a lot of space, and is
generally not adopted.

Instead, our approach places signs at strategic locations according to
human vision and navigation properties which are evaluated via
agent-based simulations. The sign placement process is performed as an
optimization against a number of cost terms reflecting the quality of
the wayfinding experience brought about by the signs. The optimization
starts with the trivial solution of placing a sign at each node along
a path, then iteratively alters the sign placement to optimize the
costs.

One advantage of this approach compared to the alternative approach of
iteratively adding signs from scratch is that the optimization process
of this approach is much more tractable, because the initial solution
and each intermediate solution represent a feasible wayfinding
solution even though they may contain redundant signs and the sign
distribution may not be ideal. As we experienced in our experiments,
this approach allows the optimizer to progress stably and
conservatively to achieve a refined sign placement solution effectively.

\begin{figure*}
 \centering
\subfloat[Initialization]{\includegraphics[width=0.25\linewidth]{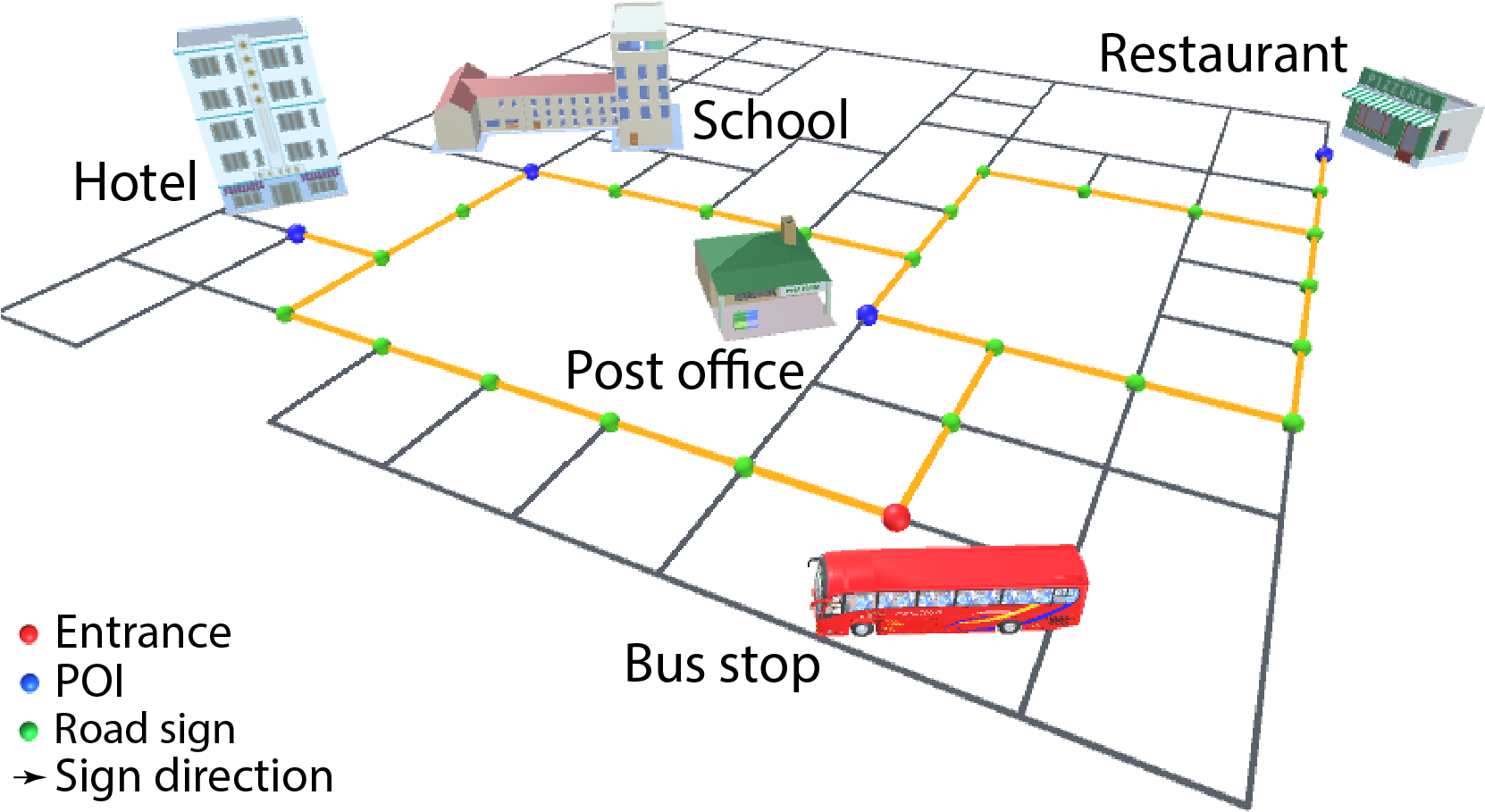}}
%\subfloat[Iteration 10]{\includegraphics[width=0.2\linewidth]{newversion/ai_rs_iterative_10}}
\subfloat[Iteration 100]{\includegraphics[width=0.25\linewidth]{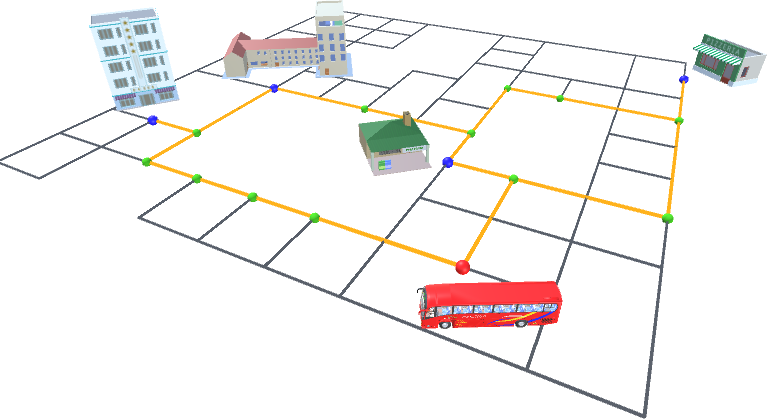}}
\subfloat[Iteration 500]{\includegraphics[width=0.25\linewidth]{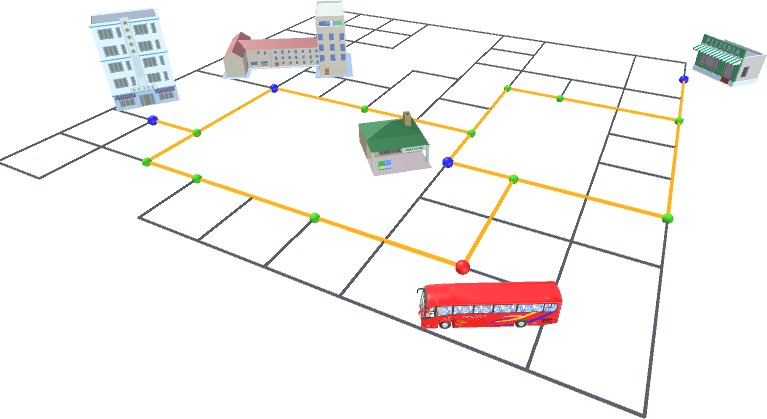}}
\subfloat[Iteration 1,000 (result)]{\includegraphics[width=0.25\linewidth]{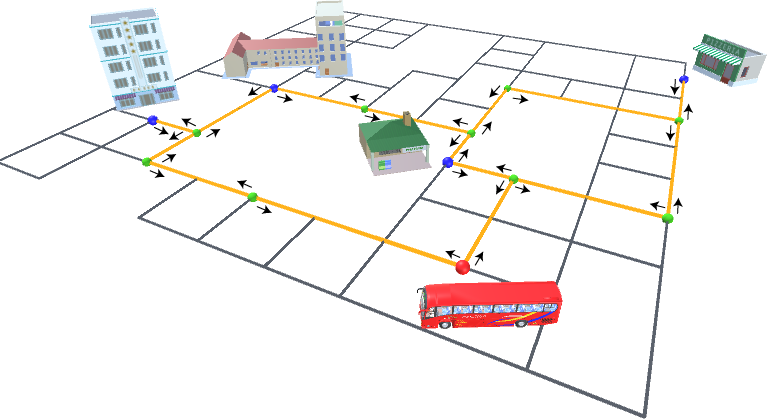}}
\\
\caption{{\footnotesize Sign placement generated over the iterations of the optimization in the agent-based sign refinement step. (a) Initialization. Road signs are placed at every intersection along the paths generated in the wayfinding scheme optimization step. (b) Iteration 100. Some redundant road signs have been removed. (c) Iteration 500. Redundant road signs along the bottom path have been further removed. (d) Result. The signs on the bottom path are combined into a single sign placed near the middle of the path. The redundant sign on the upper path has been removed.}}
 \label{fig:rs_iterations}
\vspace{-4mm}
\end{figure*}

\subsection{Representation}
A sign placement solution refers to placing signs at certain nodes of
the input layout. Given the path of each source-destination pair
(computed from Section~\ref{sec:optimization}), a good sign placement
solution guides each pedestrian to walk from the source to the
destination through the path effectively. Note that there could exist
multiple reasonable sign placement solutions. The goal of this step is
to locate one of such solutions through an optimization.

In case a road connecting two adjacent nodes is long, we may want to
place signs at some intermediate locations along the road to reassure
the pedestrian about his walking direction. Therefore, for roads
longer than a distance threshold $d_{\textrm{r}}$, our approach adds extra
nodes between the two end nodes of the road in a pre-processing step,
such that the distance between any two adjacent nodes is shorter than
$d_{\textrm{r}}$. These extra nodes serve as additional potential
locations for placing signs. $d_{\textrm{r}}$ can be empirically set by
the designer depending on how frequently a pedestrian should be
reassured about his direction. For example, for a subway station, the
designer can use a smaller $d_{\textrm{r}}$ such that more signs will be
generated along a long road to reassure pedestrians that they are
walking towards a desired destination (\eg, a platform). For our
illustration example, \emph{City}, we set $d_{\textrm{r}}$ to be 50 meter.

More specifically, given the graph $G=\{V, E\}$ representing the input
layout, we extend $V$ to $V'$ to include the extra nodes added. A sign
placement solution is represented by $S=\{(v_i,\phi_i)\}$, where
$v_i \in V'$ is the node at which sign $\phi_i$ is placed. $\phi_i$
contains the sign's attributes such as its arrow direction and the name
of the destination it is referring to. Our optimization searches for
a desirable sign placement solution $S^*$ by minimizing a total cost
function $C^{\textrm{S}}_{\textrm{all}}(S)$:

\begin{equation}
C^{\textrm{S}}_{\textrm{all}}(S) = w^{\textrm{N}}_{\textrm{sign}} C^{\textrm{N}}_{\textrm{sign}} + w^{\textrm{D}}_{\textrm{sign}} C^{\textrm{D}}_{\textrm{sign}} + w^{\textrm{F}}_{\textrm{sign}} C^{\textrm{F}}_{\textrm{sign}},
\end{equation}

\noindent where $C^{\textrm{N}}_{\textrm{sign}}$ and $C^{\textrm{D}}_{\textrm{sign}}$ are regularization costs; $C^{\textrm{F}}_{\textrm{sign}}$ is the agent-based simulation cost for estimating the wayfinding failure induced by the sign placement solution $S$. $w^{\textrm{N}}_{\textrm{sign}}$, $w^{\textrm{D}}_{\textrm{sign}}$ and $w^{\textrm{F}}_{\textrm{sign}}$ are the weights of the cost terms, which are respectively set as $1$, $1$ and $10$ by default.

\subsection{Sign Placement Cost Terms}
\label{sec:signcosts}

\ssection{Number Of Signs.} We include a cost term to regularize the number of signs in the sign placement solution, to penalize the existence of redundant signs:

\begin{equation}
C^{\textrm{N}}_{\textrm{sign}}(S) = \frac{N(S)}{|V'|},
\end{equation}

\noindent where $N(S)$ is the number of placed signs; $|V'|$ being the total number of nodes (\ie, potential locations for placing signs) is a normalization constant.

\ssection{Distribution of Signs.} In a real world design, signs are often evenly distributed along a path, which serve the purpose of regularly reassuring a pedestrian about his direction towards the destination. Accordingly, we include a cost term to regularize the distribution of signs:

\begin{equation}
C^{\textrm{D}}_{\textrm{sign}}(S) = \frac{1}{|P|} \sum_{p \in P} \frac{\sigma(p)}{L(p)},
\end{equation}

\noindent where $|P|$ is the number of source-destination pairs; $\sigma(p)$ is the standard deviation of the distances between any two adjacency signs on path $p$, and $L(p)$ is the length of path $p$.

%The $SD(p)$ function is 

%\begin{equation}
%C^{\textrm{}}_{\textrm{SD}}(p) = \sqrt{\frac{1}{|S|} \sum_{i=0}^{|S|-1} (x_i - \overline{x})^2}.
%\end{equation}

\ssection{Wayfinding Failure.} The placed signs should effectively guide the pedestrians from the sources to the destinations. We include a cost term to penalize wayfinding failure:

% \begin{equation}
% C^{\textrm{WF}}_{\textrm{sign}}(s) = \frac{F(s)}{A},
% \end{equation}
 
\begin{equation}
  C^{\textrm{F}}_{\textrm{sign}}(S) = \left\{
  \begin{array}{ll}
    F(S) & \text{if } F(S) \leq \mu,\\
    +\infty & \text{otherwise },
     \end{array}
  \right.
\end{equation}

\noindent where $F(S)$ is the percentage of agents who fail to reach their destinations under the current sign placement $S$. $F(S)$ is obtained by performing an agent-based simulation with sign placement $S$. $\mu$ is a failure tolerance level specified by the designer, which is set as $20\%$ by default.

\subsection{Agent-based Evaluation}
In each iteration of the optimization, we employ an agent-based
simulation to evaluate the wayfinding experience under the current
sign placement $S$, to obtain $F(S)$ used for computing the wayfinding
failure cost.

\ssection{Agent Model.} Each agent mimics a pedestrian walking from a
source to a destination. We model each agent with wayfinding behavior
according to Montello and Sas~\cite{hfwn}. The agent starts from the
source. It can see any unoccluded sign within visible distance
$d_{\textrm{v}}$. Whenever it sees a sign pointing to its destination,
it will follow the sign to choose a direction to walk. If it arrives
at an intersection but is unsure about which road to take out of
several roads connected to that intersection, it will randomly choose
a road to walk with equal probability. To more realistically model
mistakes that humans can make throughout a navigation, each agent has
a probability $\Pr_{\textrm{miss}}$ of missing a sign even within
sight.

\ssection{Simulation.} For each source-destination pair, $100$ agents
are employed to walk from the source to the destination using the
agent model described. At the end of the simulation, we count the
number of agents that can successfully reach their destinations, and
hence compute $F(S)$.

A ``success'' is defined as follows: let $d_{\textrm{b}}$ be the
``baseline'' walking distance from the source to the destination if no
mistake is made (\ie, $\Pr_{\textrm{miss}}=0$) under full sign placement (\ie, a sign is
placed at every node along a path). If an agent, given the chances of
making mistakes and under the current sign placement $S$, can walk from
the source to the destination by a distance no longer than
$\lambda d_{\textrm{b}}$, the navigation is considered as a
success. The navigation is counted as a failure otherwise. We use
$\lambda = 1.5$ in our experiments.

% If an agent walks longer than distance $\alpha d_{\textrm{base}}$ but still cannot
% reach the destination, the navigation ends and is automatically
% counted as a failure. We count the resulting sign placement (after the
% tentative removal of a sign) as failure if $\beta\%$ of the agents
% fail to reach
% % acceptable if $\beta\%$ of the agents can successfully reach
%  their destinations, and hence we confirm the
% removal of the sign randomly selected in this iteration. Otherwise we
% add the sign back. We set the threshold $\beta$ as $0.1$ in our
% experiments. 

\subsection{Sign Refinement by Optimization}
\ssection{Initialization.} Our optimization is initialized with the
full sign placement solution, \ie, a sign is placed at every node
along the path from the source to the destination of each
source-destination pair. Although this sign placement can lead the
pedestrians to their destinations, it consists of a lot of redundant
signs that could be removed without affecting the pedestrians' ability
to find their ways. We apply a stochastic, agent-based optimization to
search for a reasonable sign placement solution.

\ssection{Iterative Refinement.} Our optimization proceeds
iteratively. At each iteration, a move is randomly proposed to alter
the sign placement solution whose quality is evaluated using the
total cost function $C^{\textrm{S}}_{\textrm{all}}(S)$. The moves include:

\begin{itemize}
\item Adding $1$ or $2$ signs to $1$ or $2$ source-destination pairs.
\item Removing $1$ or $2$ signs from $1$ or $2$ pairs.
\item Moving a sign from one node to another node of a source-destination pair.
\end{itemize}

The proposed solution is accepted with an acceptance probability
determined by the Metropolis criterion as described by
Equation~\ref{eqn:accept_prob}, using
$C^{\textrm{S}}_{\textrm{all}}(S)$ as the cost function. The
optimization terminates if the absolute change in cost is less than
$1\%$ over $50$ iterations.

Figure~\ref{fig:rs_iterations} shows the sign placement over
iterations for the illustrative example. In this example, the
source-destination pairs include walking from the \emph{Bus Stop} to
each POI, and walking between every pair of POIs. Each iteration of
the optimization takes about 0.01 second to finish in our experiments.
It takes about $1,000$ iterations (about $10$ seconds) to finish the
sign placement optimization for this example.

\section{Experiments and Results}
\label{sec:results}
% \subsection{Implementation}

We implemented our approach as a plugin for the Unity 5 game engine
using C\#, which level designers can use to create a wayfinding scheme
of a given layout. We run our experiments using a Macintosh machine
equipped with a $2.3$ GHz Intel Core i$7$ processor and $8$GB of
RAM. Generating a wayfinding scheme for a layout similar to the
illustrative example, \emph{City}, takes about $40$ seconds using our
current implementation.

\subsection{Different Layouts}
\label{sec:layouts}

\begin{figure*}[htp]
\begin{center}
\begin{tabular}
{
@{\hspace{0mm}}c@{\hspace{2mm}}c@{\hspace{2mm}}c@{\hspace{2mm}}c
@{\hspace{2mm}}c@{\hspace{2mm}}c@{\hspace{2mm}}c@{\hspace{2mm}}c
@{\hspace{2mm}}c
}
\includegraphics[width=0.41\linewidth]{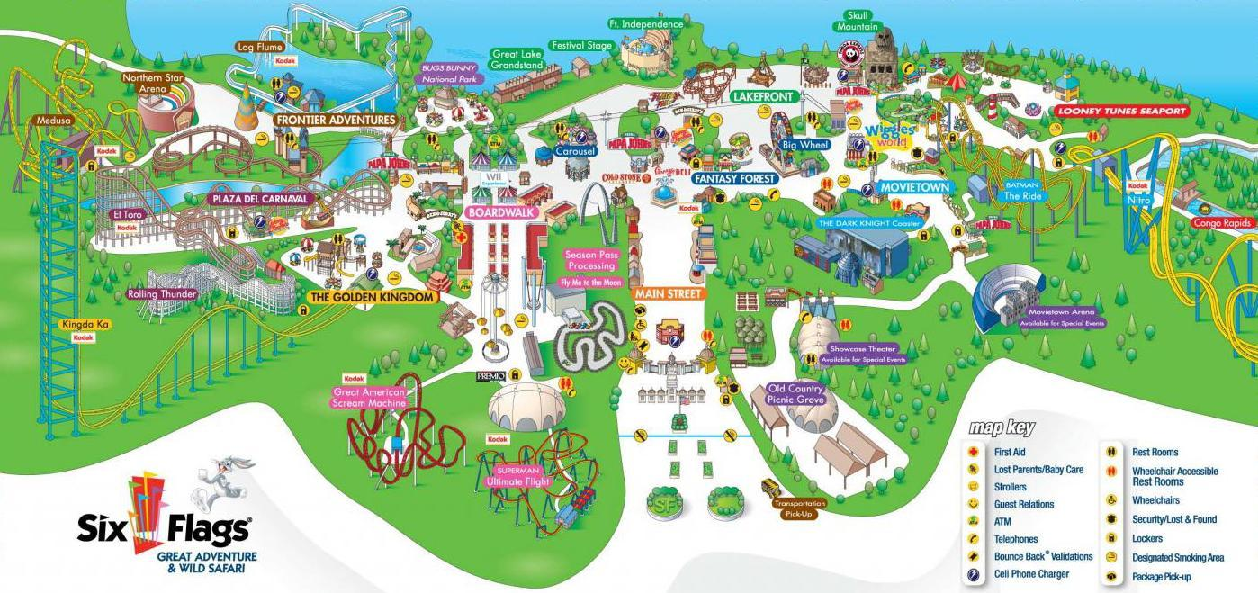} &
\includegraphics[width=0.29\linewidth]{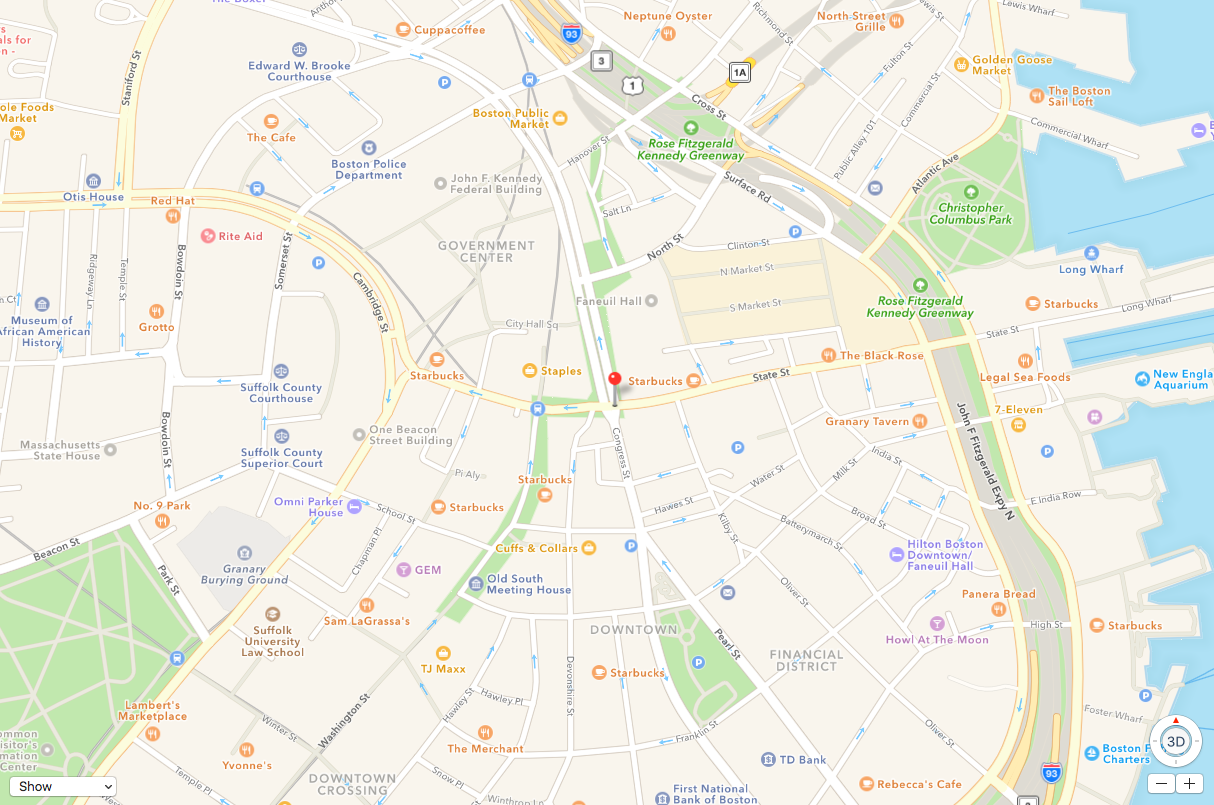} &
\includegraphics[width=0.28\linewidth]{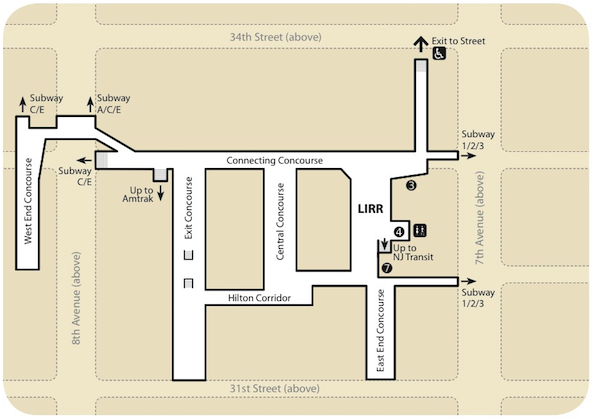} \\
{\footnotesize Amusement Park} & 
{\footnotesize Downtown} & 
{\footnotesize Penn Station} 
\end{tabular}
\end{center}
\caption{{\footnotesize Maps and 3D models from which the layouts are extracted.}}
\label{fig:result:placement_input_real}
\vspace{-5mm}
\end{figure*}

\begin{figure*}[htp]

\begin{center}
\begin{tabular}
{
@{\hspace{0mm}}c@{\hspace{4mm}}c@{\hspace{4mm}}c@{\hspace{4mm}}c
@{\hspace{4mm}}c@{\hspace{4mm}}c@{\hspace{4mm}}c@{\hspace{4mm}}c
@{\hspace{4mm}}c
}

\begin{sideways}\parbox{60mm}{\centering\footnotesize\em Amusement Park}\end{sideways} &
\includegraphics[width=0.48\linewidth]{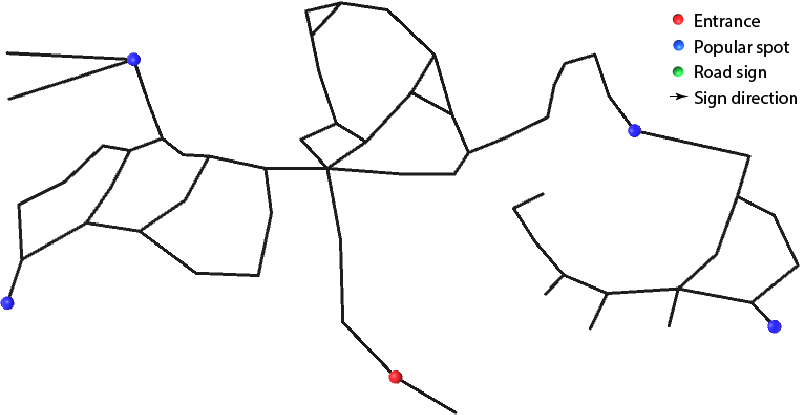} &
\includegraphics[width=0.48\linewidth]{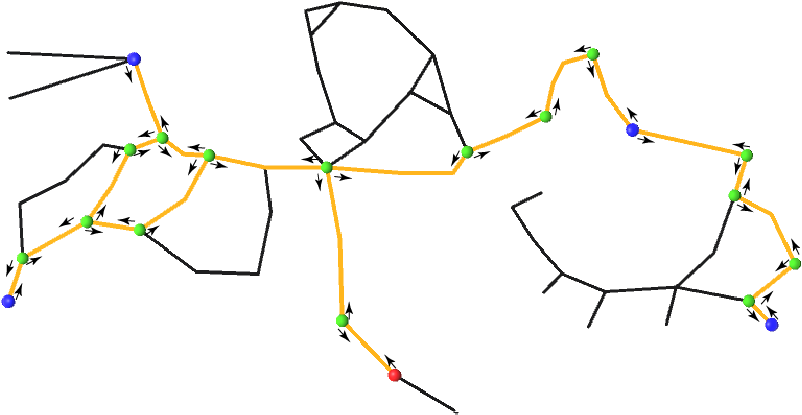} \\
%\vspace{-3mm}
\begin{sideways}\parbox{60mm}{\centering\footnotesize\em Downtown}\end{sideways} &
\includegraphics[width=0.48\linewidth]{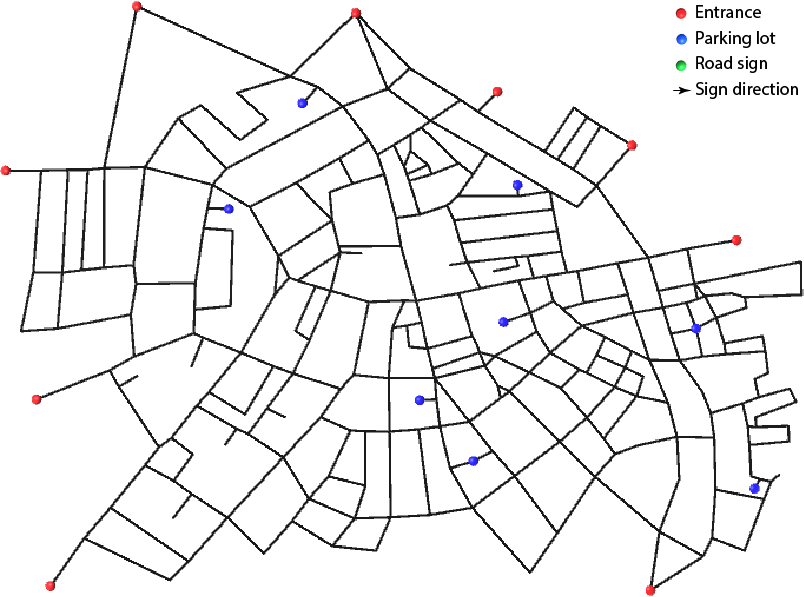} &
\includegraphics[width=0.48\linewidth]{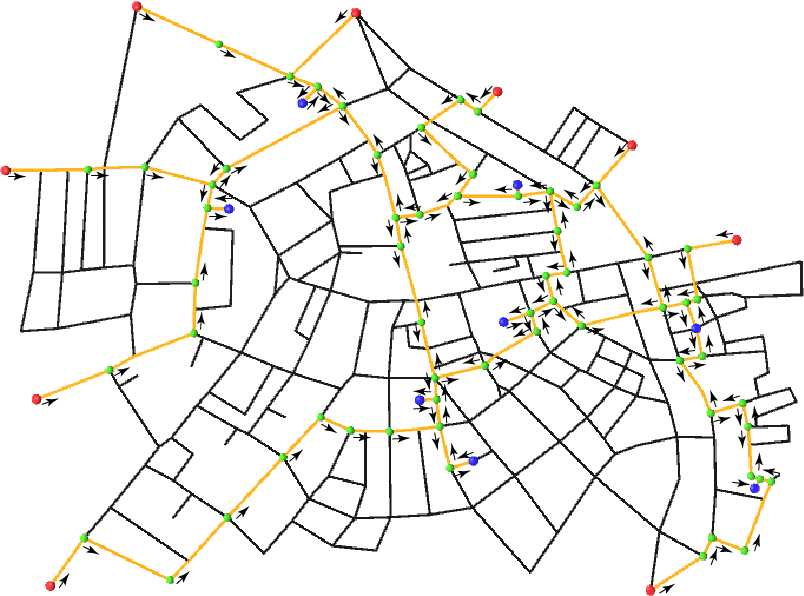} \\
%\vspace{-3mm}
\begin{sideways}\parbox{60mm}{\centering\footnotesize\em Penn Station}\end{sideways} &
\includegraphics[width=0.48\linewidth]{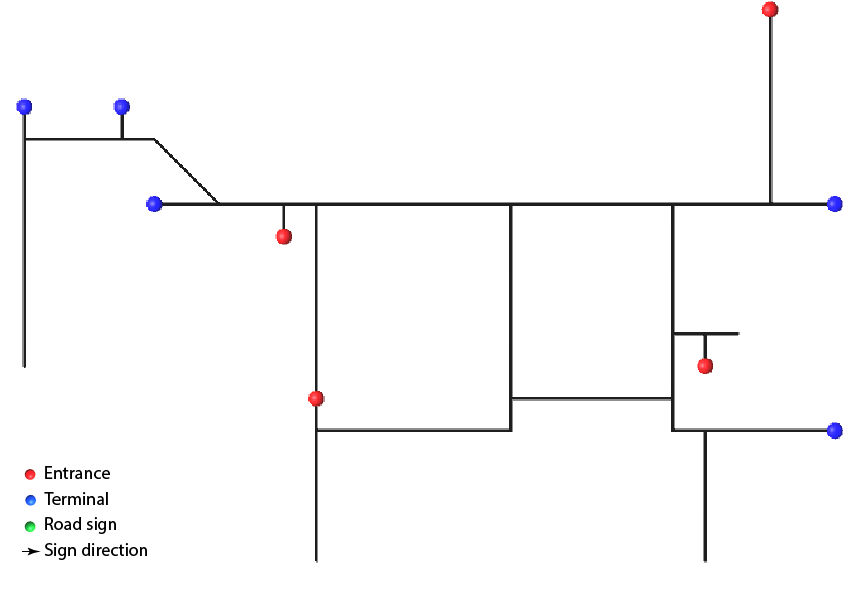} &
\includegraphics[width=0.48\linewidth]{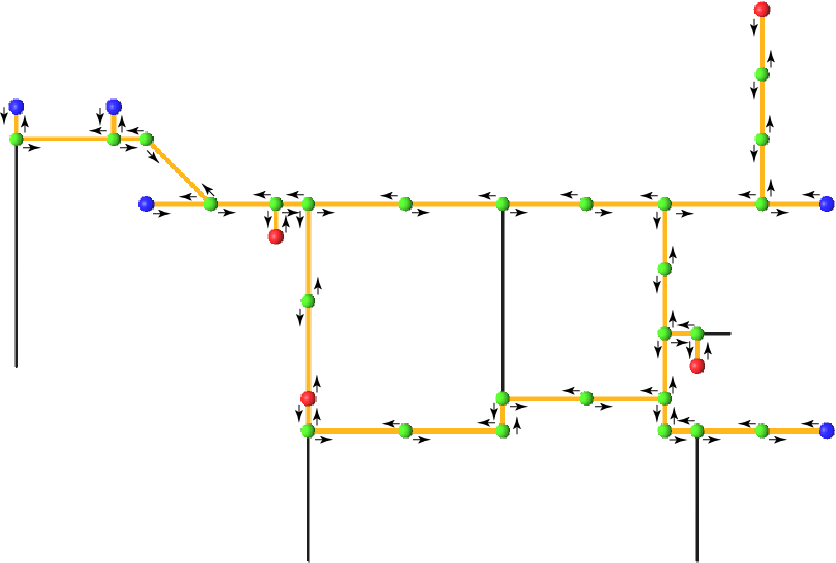} \\
%\vspace{-3mm}
%{} & {\footnotesize 1} & {\footnotesize 2} \\
\end{tabular}

\caption{{\footnotesize Wayfinding designs generated for \emph{Amusement Park}, \emph{Downtown} and \emph{Penn Station}.}}
\label{fig:result:placement1}
\end{center}  %all terms already italic in caption
\end{figure*}

We used our approach to generate wayfinding designs for different
layouts: \emph{Amusement Park}, \emph{Downtown} and \emph{Penn
  Station}. Figure~\ref{fig:result:placement_input_real} shows the
maps from which the layouts are extracted following the procedure in
Section~\ref{sec:representation}. Figure~\ref{fig:result:placement1}
shows the wayfinding designs generated by our approach. We describe
the details of each generation in the following. Please also refer to
the supplementary material for details of the generated wayfinding
schemes of the layouts, and also for the results of two more layouts:
\emph{City} and \emph{Canyon}, which demonstrate how our approach could be
applied to generate wayfinding designs for 3D virtual environments and
with robustness as a key consideration.

\ssection{Amusement Park.} We use the layout of an amusement park, Six
Flags New England, as input (see Figure~\ref{fig:result:placement1}). The
POIs, in this case, represent the popular spots the visitors would
like to visit. The source-destination pairs involve all pairs of
entrances and popular spots, and all pairs of popular spots. In
addition to the popular spots, we expect there are street performances
and stalls in the park, which might distract visitors from navigating
to their destinations. To model such distractions, we set the missing
chance $\Pr_{\textrm{miss}}$ in the agent simulation of the sign
placement optimization step to a relatively high level of
$0.2$. Besides, we assume that visitors highly prefer to walk shorter
and more direct paths to their destinations if possible, therefore we
use larger values of $5$ for the weights
$w^{\textrm{L}}_{\textrm{local}}$ and
$w^{\textrm{N}}_{\textrm{local}}$ of the local path length and local
path node costs.

Figure~\ref{fig:result:placement1} shows the generated design. Our
approach generates a path for each source-destination pair. It places
road signs densely at each intersection along the paths to ensure the
robustness of the wayfinding system. The left-hand side of the layout
shows a shortcut generated which is part of the paths from the popular
spot at the lower left to the popular spots on the right. The shortcut
allows visitors to walk shorter paths to their destinations, and is
also more direct for the visitors as it passes through fewer
intersections ($2$ instead of $3$) compared to the alternative path
above.

% \begin{figure*}[htp]
% \begin{center}
% \begin{tabular}
% {
% @{\hspace{0mm}}c@{\hspace{2mm}}c@{\hspace{2mm}}c@{\hspace{2mm}}c
% @{\hspace{2mm}}c@{\hspace{2mm}}c@{\hspace{2mm}}c@{\hspace{2mm}}c
% @{\hspace{2mm}}c
% }
% \includegraphics[width=0.25\linewidth]{newversion/sixflag3} &
% \includegraphics[width=0.18\linewidth]{newversion/Boston_Parking_Lot} &
% \includegraphics[width=0.17\linewidth]{newversion/penn_station_lower} &
% \includegraphics[width=0.18\linewidth]{newversion/city3D2} &
% \includegraphics[width=0.18\linewidth]{newversion/GrandCanyon} \\
% {\footnotesize Amusement Park} & 
% {\footnotesize Downtown} & 
% {\footnotesize Penn Station} & 
% {\footnotesize City} & 
% {\footnotesize Canyon}
% \end{tabular}
% \end{center}
% \caption{{\footnotesize Maps and 3D models from which the layouts are extracted.}}
% \label{fig:result:placement_input_real}
% \vspace{-5mm}
% \end{figure*}

\begin{figure*}
\centering
\subfloat[Default Parameters]{\includegraphics[width=0.25\linewidth]{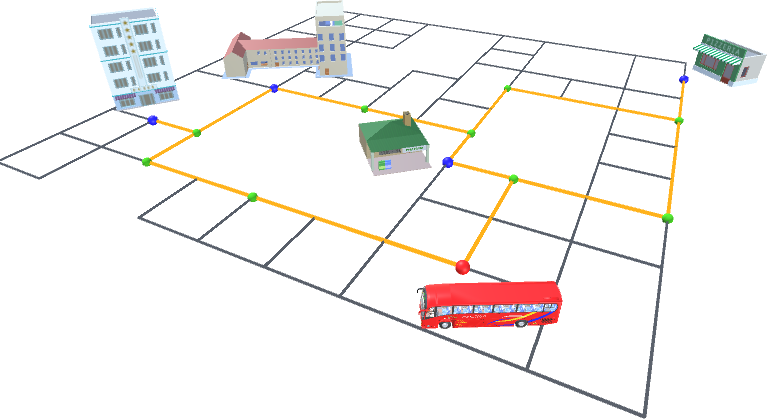}}
\subfloat[High Miss Chance]{\includegraphics[width=0.25\linewidth]{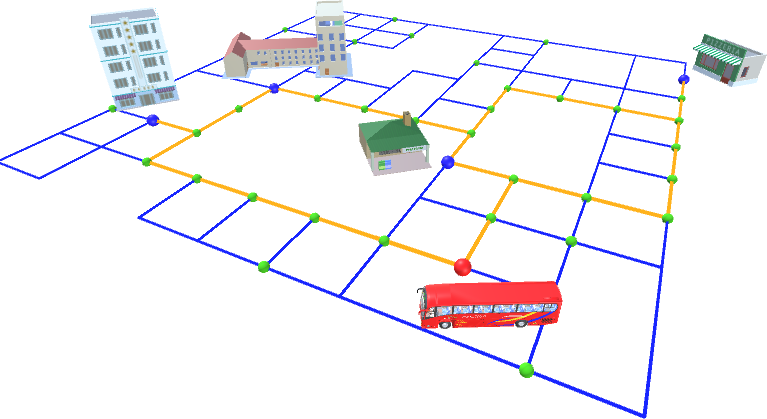}}
\subfloat[Low Wayfinding Failure Weight]{\includegraphics[width=0.25\linewidth]{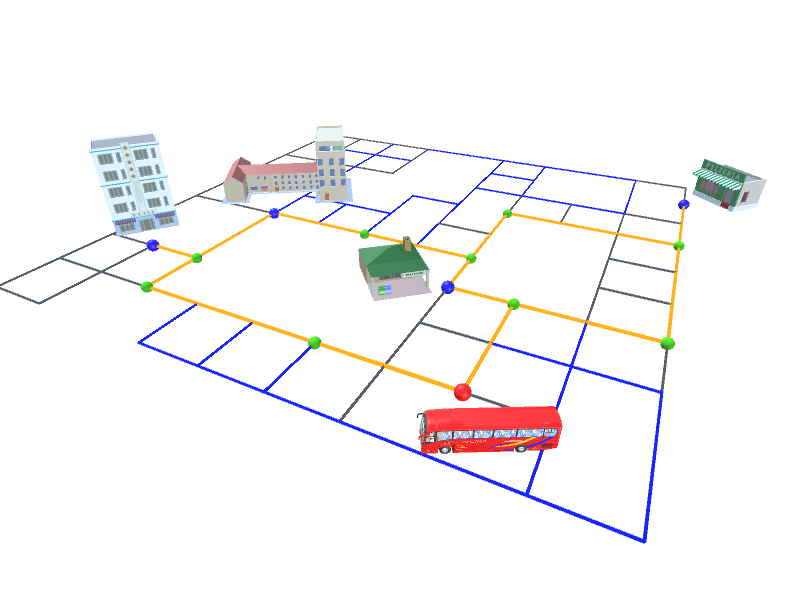}}
\subfloat[High Visibility]{\includegraphics[width=0.25\linewidth]{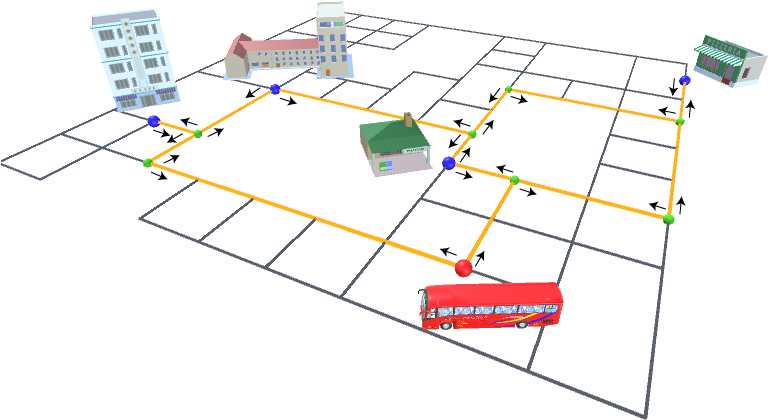}}
\caption{{\footnotesize 
Effects of changing the parameters of the agent-based sign placement step. (a) Default parameters. Roads that belong to a path of the wayfinding scheme are shown in yellow. (b) High miss chance. The blue lines indicate the roads that the pedestrians have walked but do not belong to any path. The generated wayfinding design is more robust against navigation mistakes. Signs are placed densely, and are also placed on some roads not belonging to any path to guide the pedestrians back to the correct paths in case they get lost. (c) Low wayfinding failure weight. Fewer signs are placed as wayfinding failure is more tolerable. (d) High visibility. Fewer signs are placed as the pedestrians can see signs far away.}}
\label{fig:parameters}
\vspace{-4mm}
\end{figure*}

\ssection{Downtown.} This example uses the layout of Downtown Boston
as input. The goal is to place road signs that guide drivers to an
available parking lot nearby. The entrances refer to the major roads
through which most cars enter the Downtown area. The POIs are the
parking lots, which are placed at the same locations as the real
parking lots found on Google map. We suppose that all the parking lots
are run by the same company, hence there are signs showing the way
from one parking lot to a nearby parking lot within $0.2$ mile, such
that if a parking lot is full the driver can follow the signs to a
nearby parking lot. Accordingly, we define the source-destination
pairs to connect each entrance to its nearest parking lot, as well as
to connect each parking lot to its nearby parking lot. The latter type
of pairs are given a relatively larger importance value
($\kappa_p=0.8$ instead of $0.5$ given to other pairs), so that our
system will prefer to find shorter paths passing through fewer
intersections for the paths that connect one parking lot to a nearby
parking lot, to help drivers to get to an alternative parking lot
more easily in case a parking lot is full.

Figure~\ref{fig:result:placement1} shows the generated design. A path
is generated to connect each entrance to its nearest parking
lot. Short and direct paths are also generated to connect parking lots
to nearby parking lots. While there are many possible paths that can
be chosen as the layout comprises a network of many streets, our
approach chooses paths which are straight and consists of few turns,
as the local path node cost penalizes the inclusion of intersections
and the local path angle cost discourages orientation changes.

\ssection{Penn Station.} This example uses the lower level of the Penn
Station as input. In this example, the entrances refer to the gates
and the stairs from the upper level. The POIs refer to the
terminals. The source-destination pairs include every pair of entrance
and terminal, and every pair of terminals (for modeling the situations
where a passenger wants to transfer from one terminal to another
terminal). As the station is expected to be crowded, the visibility
$d_{\textrm{v}}$ of the agents is set to a relatively low value of
$10$ meters to account for the occlusion by human crowd, and the miss
chance $\Pr_{\textrm{miss}}$ is set to a relatively high level of
$0.2$. Figure~\ref{fig:result:placement1} shows the generated
wayfinding design. The road signs are placed densely and are also
placed at non-intersection nodes, to counteract the higher miss chance
by reassuring pedestrians about their directions.

\subsection{Changing Agent Parameters}
\label{sec:changing}
We further experimented with changing the parameters of the
agent-based sign placement process using the \emph{City} layout. In
the default settings, the missing chance $\Pr_{\textrm{miss}}$ is set
to $0\%$, the weight $w^{\textrm{F}}_{\textrm{sign}}$ of the
wayfinding failure cost term is set to $10$ and the visibility
distance $d_\textrm{d}$ is set to $125$
meters. Figure~\ref{fig:parameters}(a) shows the resulting sign
placement generated with the default parameters.

We experimented with increasing the missing chance
$\Pr_{\textrm{miss}}$ to $10\%$. Figure~\ref{fig:parameters}(b) shows
the resulting sign placement. Our system places more signs so as to
increase the robustness of the wayfinding design against navigation
mistakes. In addition, some signs are placed on the roads not
belonging to any path for guiding the agents back to the correct
paths. 

Next, we experimented with lowering the weight of the wayfinding
failure cost term to $0.01$. Figure~\ref{fig:parameters}(c) shows the
resulting sign placement. Our system keeps fewer signs, because it is
acceptable even if some agents make mistakes and do not walk to the
destination within a desired period of walking. The pedestrians walk
along some roads (shown in blue) not belonging to any path. This
setting maybe useful for some situation where it is not critical for
the agents to reach the destination, and when space is better
preserved for other uses. For example, in a flea market, it may not be
critical for the pedestrians to visit each stall as they are expected
to wander around in the market.

Finally, we experimented with increasing the visibility to $250$
meters. Figure~\ref{fig:parameters}(d) shows the result. Our system
keeps only a fewer signs because the pedestrians are capable of seeing
signs at a farther distance. This setting is useful for modeling
situations where the signs are big (such as those shown in billboards)
and can be seen far away.

\subsection{Visualization}
\label{sec:vis}
\begin{figure}[h!]
\centering
\subfloat[{\footnotesize Accessibility of \emph{Post Office}}]{\includegraphics[width=0.49\linewidth]{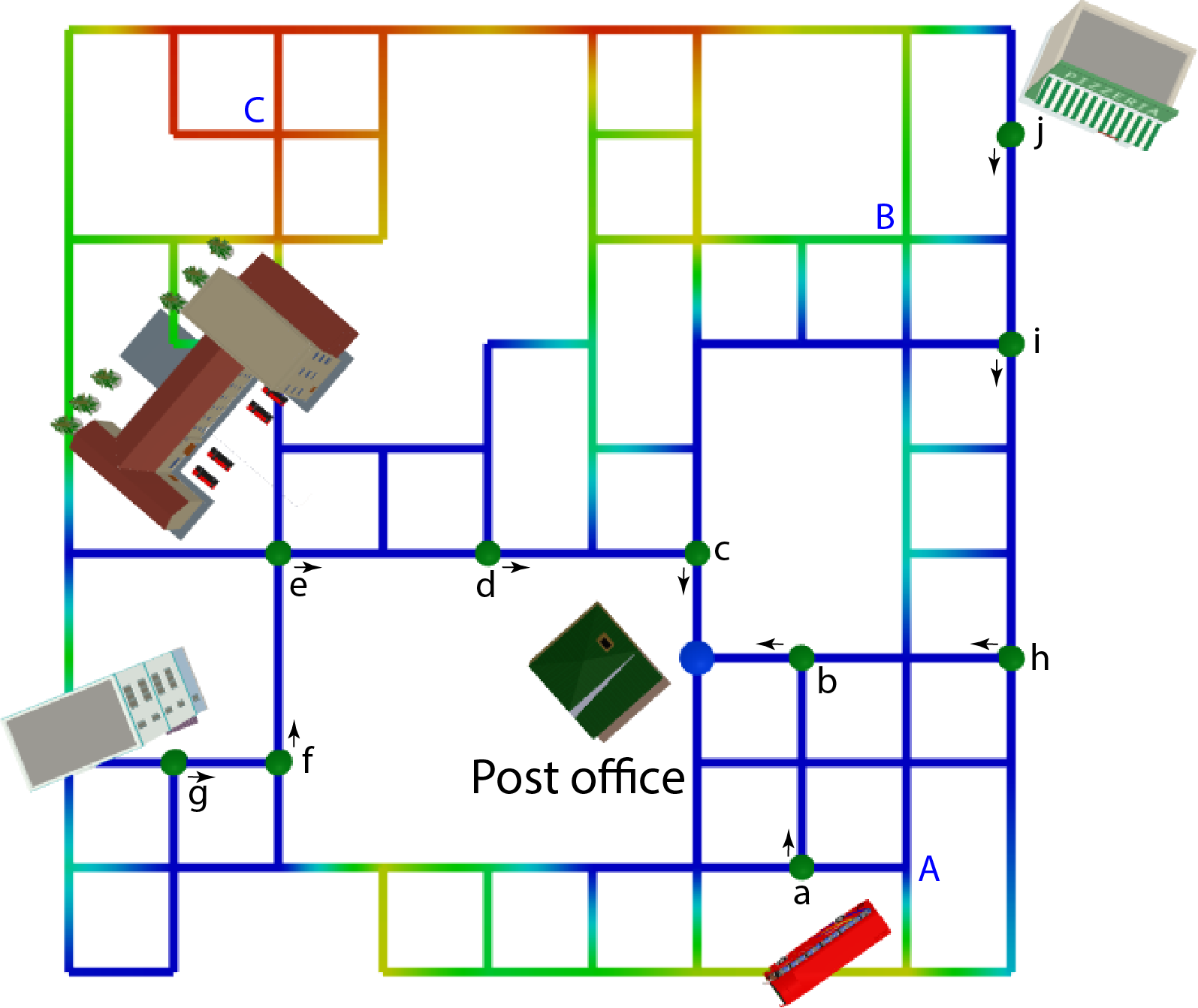}}
\subfloat[{\footnotesize Remove Blind Zone (at C)}]{\includegraphics[width=0.49\linewidth]{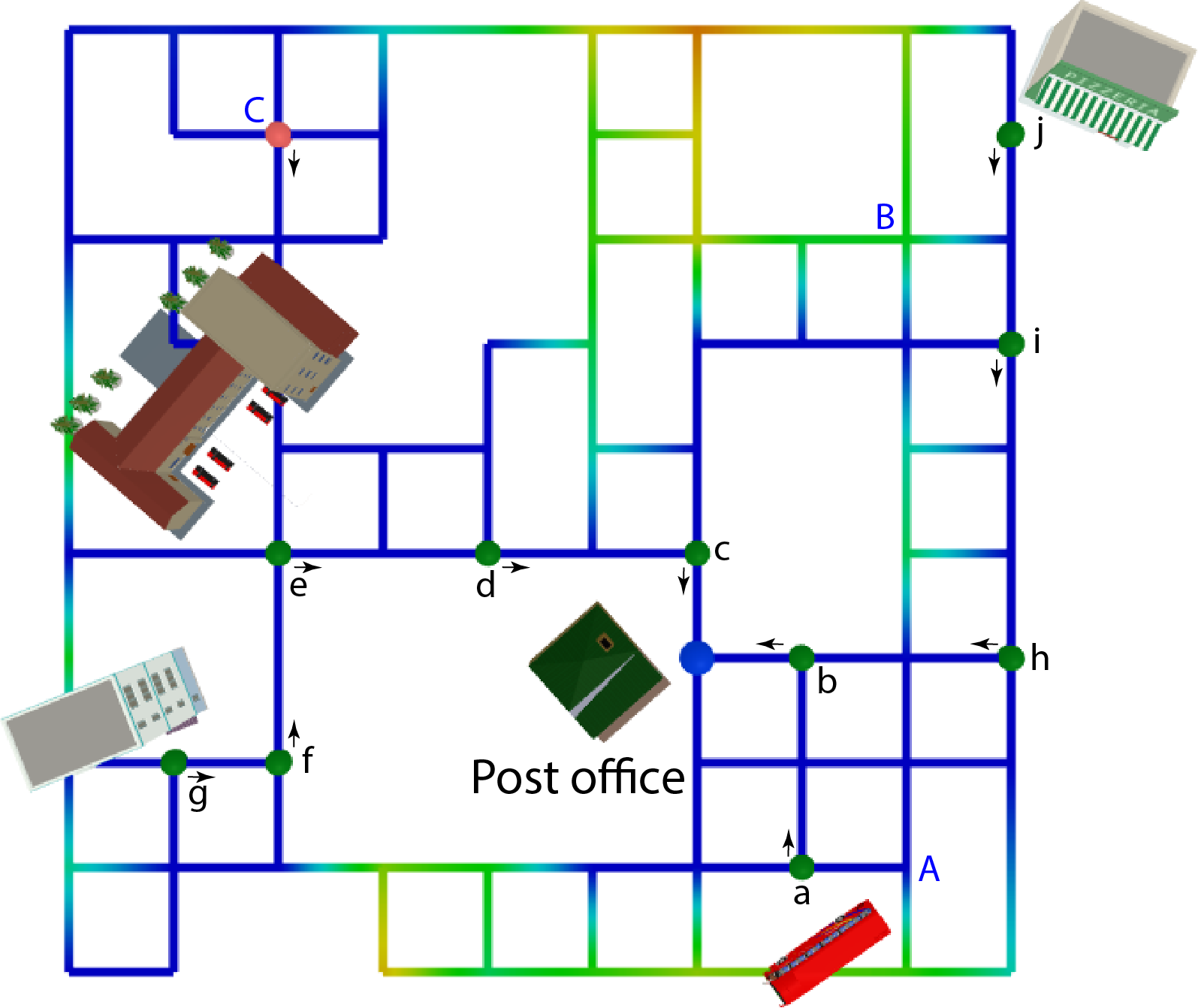}}
\\
{\includegraphics[width=0.85\linewidth]{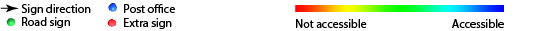}}
\caption{{\footnotesize (a) Accessibility of the \emph{Post Office}
    visualized as a heatmap. The accessibility from location A is good
    as depicted by the blue color, because a pedestrian can easily see
    and follow a sign on the west to walk to the \emph{Post
      Office}. The accessibility from location B is fair as depicted
    by the green color. A pedestrian starting at B has a $50\%$ chance
    to walk to the street towards the north or the east and get
    lost. The accessibility from location C is poor as depicted by the
    red color. Without any sign nearby, a pedestrian starting at C has
    a very low chance to reach the \emph{Post Office}. (b) Removing
    the blind zone at C. The user triggers our system to place a road
    sign at C showing the direction to the \emph{Post Office}. The
    accessibility around C is improved, as depicted by the change in
    color from red to blue on the updated heatmap.}}
 \label{fig:heatmap}
\end{figure}

\ssection{Destination Accessibility.} Our approach also allows the
designer to visualize the accessibility of a destination under the
generated wayfinding design. This is a very useful functionality that
can help the designer to create a wayfinding design that guides
pedestrians from different locations to walk to a destination as
desired. Figure~\ref{fig:heatmap}(a) depicts this functionality. The
accessibility of a destination (the \emph{Post Office}) is visualized
as a heatmap. Agents in the blue region can travel to the \emph{Post
  Office} successfully by following the wayfinding signs under the
current wayfinding design; while those in the red region have a low
chance of success.

To compute the accessibility heatmap with respect to a destination
specified by the designer, our system sample points at regular
intervals along all the edges of the input layout (whether the edges
are part of the paths of the generated wayfinding design or
not). Agents are employed to walk from each sample point to the
destination, in a similar fashion as in the agent-based sign placement
step (Section~\ref{sec:agent}). The rates of success are used to set
the heatmap values at that sample points; the heatmap values between
two sample points are interpolated.

Note that a destination typically does not need to be accessible from
every region, because enforcing such full accessibility will likely
involve placing a lot of signs even at some ``unimportant''
regions. For example, it may not be important to place signs to guide
pedestrians how to walk from a post office to a restaurant. By
visualizing the accessibility to a destination using a heatmap, the
designer can intuitively tell what regions are covered by the current
wayfinding design and if any improvement is needed.

\ssection{Removing Blind Zones.} If the designer wants to remove a
``blind zone'' (\ie, a region shown in red indicating low
accessibility to the destination), he can easily do so by clicking on
the red region via our user interface. Our system will automatically
place signs which guide pedestrians to walk from the clicked point to
the path leading to the destination. Agent-based evaluations will be
re-run at each sample point to update the heatmap accordingly, which
takes about $1$ second for the \emph{City}
example. Figure~\ref{fig:heatmap}(b) shows an example of removing a
blind zone.

\section{Evaluation}
\label{sec:evaluation}

\subsection{User Study} 
\ssection{Conditions.} We conducted a user study to evaluate the
effectiveness of the wayfinding designs generated by our approach. Our
user study was conducted in the \emph{City} layout used as the
illustrative example. Participants were asked to navigate from a starting
point to a destination under $4$ different wayfinding conditions:

\begin{enumerate}
\item No sign.
\item Mini-map. A mini-map that functions like a mini-map in a common first-person 3D video game is shown;
\item Full signs. In this case, we only run the wayfinding scheme optimization step to generate the paths for the source-destination pairs. Signs are placed at every node along each path.
\item Refined signs. In this case, we run the wayfinding scheme
  optimization step to generate the paths, and then the agent-based
  sign refinement step to refine the sign placement. Signs are placed
  strategically at some of the nodes along each path.
\end{enumerate}

\begin{figure}[t]
\centering
\subfloat[{\footnotesize Mini-map}]{\includegraphics[width=0.49\linewidth]{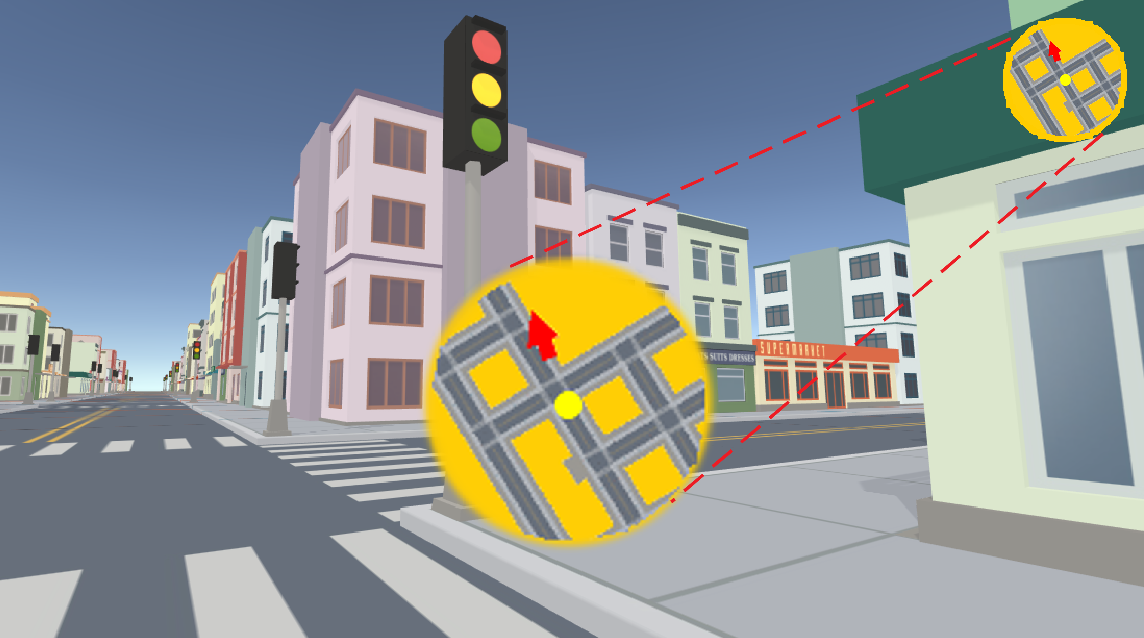}}\hfill
\subfloat[{\footnotesize Refined signs}]{\includegraphics[width=0.49\linewidth]{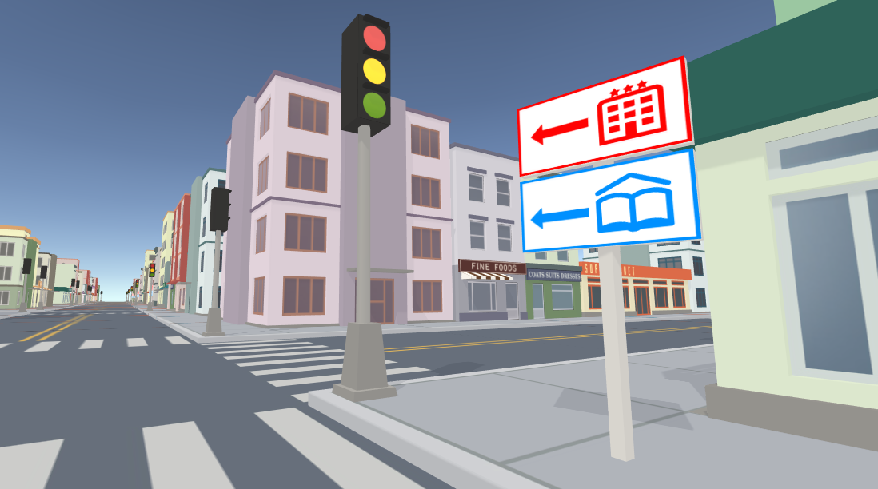}}
\\
\caption{{\footnotesize Screenshots of the user study tests under the (a) mini-map and (b) refined signs conditions.}}
 \label{fig:us}
\vspace{-5mm}
\end{figure}

Figure~\ref{fig:us} shows two screenshots of the user study tests
under the mini-map and refined sign conditions. There are $2$
different scenarios. In the first scenario, the participant was asked
to walk from the \emph{Bus Stop} to the \emph{Restaurant}. In the
second scenario, the participant was asked to walk from the \emph{Bus
  Stop} to the \emph{School}. Each scenario was tested by $80$
participants under the $4$ different wayfinding conditions (\ie, $20$
participants for each condition).

\ssection{Participants.} In total, we recruited $160$ participants through
social networks. The participants are university students. All of them have
experience with 3D video games and are familiar with the movement
control of common first-person-shooting games, which our user study
program similarly adopts. Before each test, a description of the task
and the movement control is shown to the participant, and the participant is allowed
to get familiar with the movement control in a warm-up session.

\ssection{Test Sessions.} The goal of the participant in each test is to walk
to the destination (\emph{Bus Stop} or \emph{Restaurant}) as fast as
he can. To make sure he is clear about the destination, a screenshot
of the destination is also shown to the participant before the user study
begins. Our program records the path, the distance walked and the time
taken by the participant. The test ends if the participant reaches the destination,
or if the time taken exceeds the time limit, which is defined as three
times the time needed to walk from the start to the destination
without any stop following the path generated by the wayfinding scheme
optimization step. The latter is considered as a failure case.

% user study figure
\begin{figure*}[ht]
\begin{center}
\begin{tabular}
{ c c c c
%@{\hspace{0mm}}c@{\hspace{0mm}}c@{\hspace{0mm}}c@{\hspace{0mm}}c
%@{\hspace{0mm}}c@{\hspace{0mm}}c@{\hspace{0mm}}c@{\hspace{0mm}}c
%@{\hspace{0mm}}c
}

%\begin{sideways}\parbox{60mm}{\centering\footnotesize\em }\end{sideways} &
\includegraphics[width=0.24\linewidth]{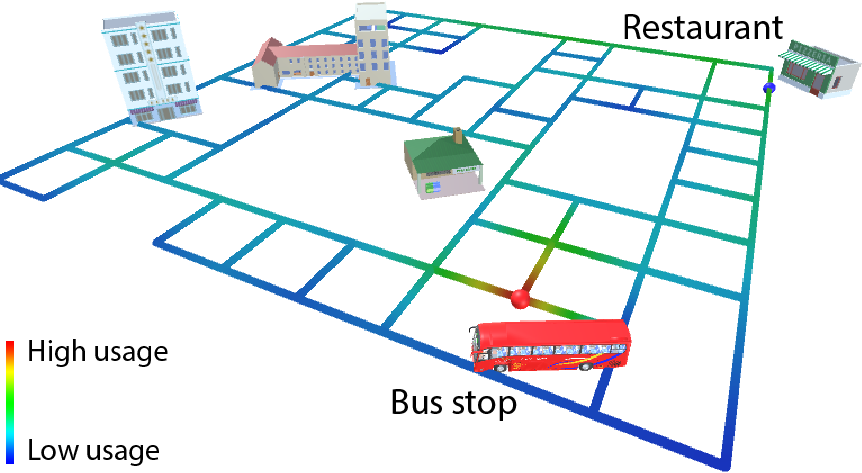} & 
\includegraphics[width=0.24\linewidth]{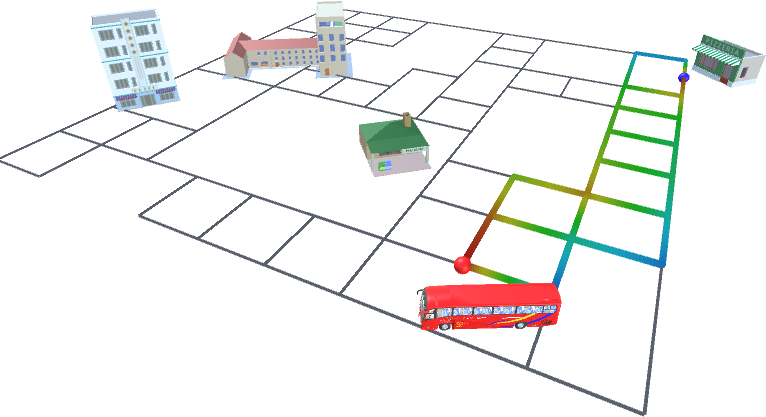} &
\includegraphics[width=0.24\linewidth]{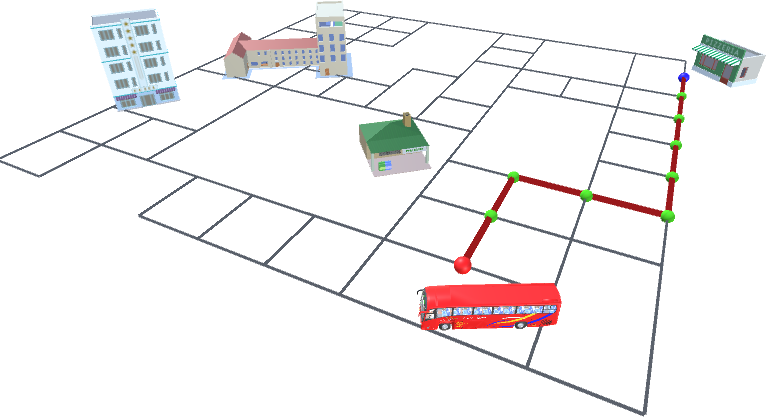} &
\includegraphics[width=0.24\linewidth]{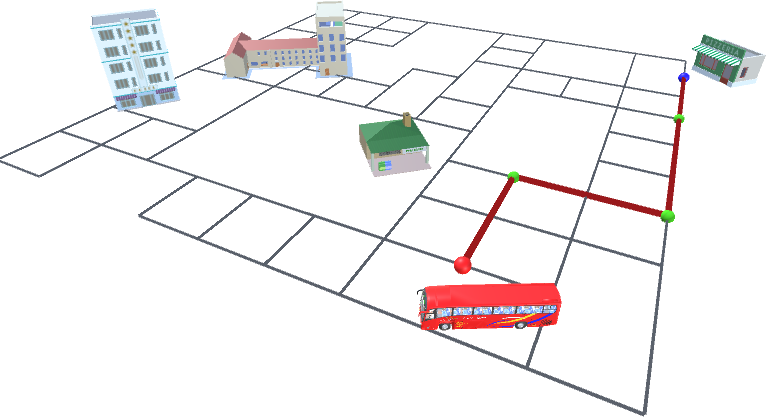} \\
{} {\footnotesize(a) No sign} & {\footnotesize(b) Mini-map} & 
{\footnotesize(c) Full signs} & {\footnotesize(d) Refined signs}\\ 
\multicolumn{4}{c}{{\footnotesize Scenario 1 (\emph{Bus Stop} to \emph{Restaurant})}}\\
\vspace{1mm}\\
%\begin{sideways}\parbox{60mm}{\centering\footnotesize\em }\end{sideways} &
\includegraphics[width=0.24\linewidth]{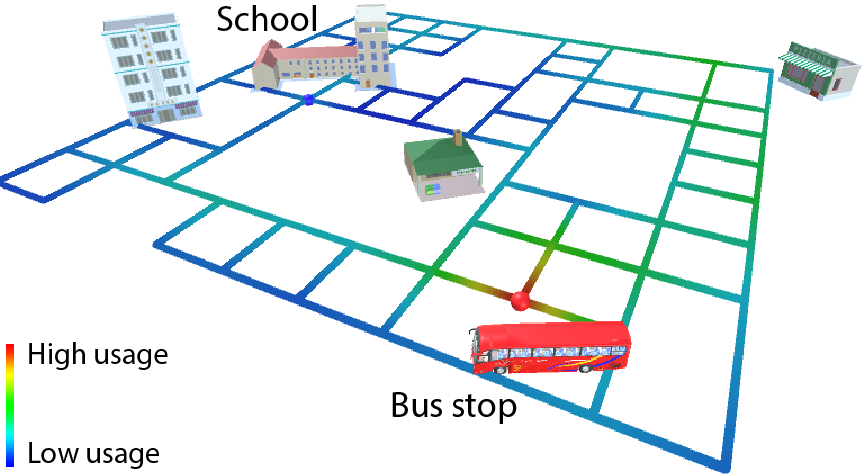} & 
\includegraphics[width=0.24\linewidth]{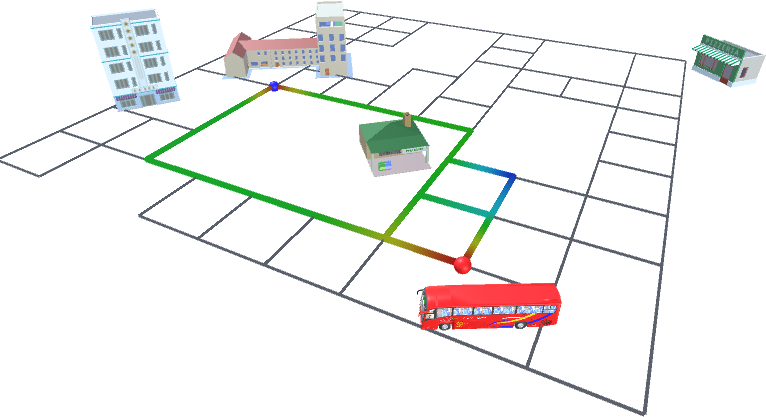} &
\includegraphics[width=0.24\linewidth]{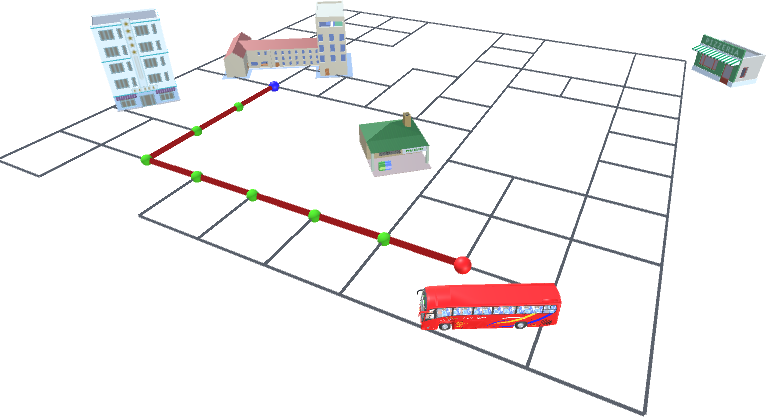} &
\includegraphics[width=0.24\linewidth]{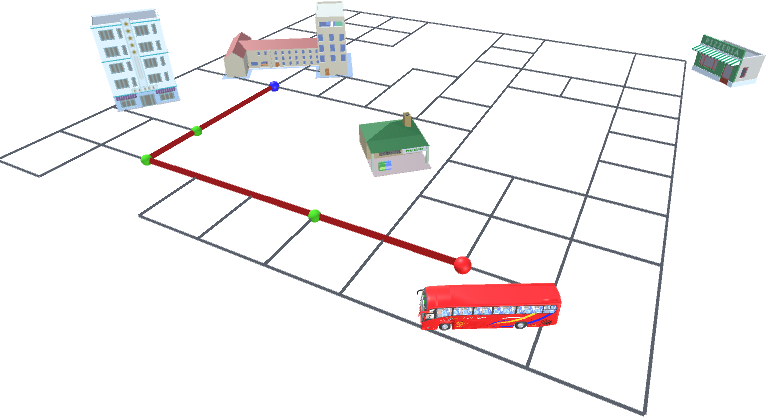} \\

{} {\footnotesize(a) No sign} & {\footnotesize(b) Mini-map} & 
{\footnotesize(c) Full signs} & {\footnotesize(d) Refined signs}\\ 
\multicolumn{4}{c}{{\footnotesize Scenario 2 (\emph{Bus Stop} to \emph{School})}}
\\

\end{tabular}
\vspace{-2mm}
\caption{
{\footnotesize 
Paths taken by the participants of the user study under different wayfinding conditions. Usage of each road is shown by its color (roads which are not walked are shown in gray). (a) With no sign, participants wandered randomly and could hardly reach the destination. (b) With a mini-map, participants walked along a similar direction towards their destinations, but the paths they took varied. (c,d) With full signs or refined signs, all participants followed the same paths to walk to their destinations. The refined sign placement is as effective as the full sign placement, but uses a significantly smaller number of signs.}}
\label{fig:result:userstudy1}
\end{center} 
\vspace{-4mm}
\end{figure*}

\subsection{Results and Analysis} 
\ssection{Path Taken.} Figure~\ref{fig:result:userstudy1} shows the
results of the user study. The paths taken by the participants are
visualized in a heatmap. The roads with high usage are shown in red,
and those with low usage are shown in blue. There are some interesting
observations. Under the no sign condition, the participants wandered
around and could barely reach the destination. Under the mini-map
condition, the participants walked towards their destinations along
similar directions. However, there are considerable variations among
the paths taken, as can be seen from the color dispersion on the
heatmaps. For example, in scenario 2, near half of the participants
took the bottom path while the other half took the upper path. Under
the full signs and refined signs conditions, all participants walked
to the destinations following the same path.

% user study table 
\begin{table}[t]
\centering

\begin{tabular}{l l l l l}
\hline
\textbf{Condition} & \textbf{Mean} & \textbf{SD} & \textbf{Success Rate} & \textbf{\#Signs}\\ [0.5ex] 
\hline
No sign & 1,128.28 & 361.22 & 55.00\% & -\\
Mini-map  & 515.54 & 35.81 & 100.00\% & -\\
Full signs  & 514.21 & 11.45 & 100.00\% & 8\\
Refined signs  & 517.83 & 11.95 & 100.00\% & 3\\
\hline
\vspace{-7pt}
\end{tabular}
\mbox{(a) Scenario 1 (\emph{Bus Stop} to \emph{Restaurant})}

\vspace{3mm}

\begin{tabular}{l l l l l}
\hline
\textbf{Condition} & \textbf{Mean} & \textbf{SD} & \textbf{Success Rate} & \textbf{\#Signs}\\ [0.5ex] 
\hline
No sign  & 1,205.67 & 187.51 & 25.00\% & -\\
Mini-map & 503.32 & 55.27 & 100.00\% & -\\
Full signs & 471.37 & 6.11 & 100.00\% & 7\\
Refined signs & 476.32 & 7.38 & 100.00\% & 3\\
\hline
\vspace{-7pt}
\end{tabular}
\mbox{(b) Scenario 2 (\emph{Bus Stop} to \emph{School})}
\label{table:tuser}
\caption{{\footnotesize Distances (in meters) walked by participants under different wayfinding conditions. For the no sign condition, only the data of the participants who successfully reached their destinations are used to compute the statistics.} }
\label{table:tuser}
\vspace{-5mm}
\end{table}

\ssection{Distance Walked.} Table~\ref{table:tuser} shows the
statistics of the distances walked by the participant under different
conditions. For the no sign condition, only the data of the
participants who could reach their destinations within the time limit
is used to calculate the statistics. For the other conditions, all
participants can reach the destinations and all data is used to
calculate the statistics.

Under the no sign condition, only $55\%$ and $25\%$ of the participants
could reach their destinations in Scenario 1 and 2 respectively. For
those who could reach the destinations, they generally needed to walk
a very long way as shown by the large mean values.

Under the mini-map condition, all participants could reach their
destinations. In Scenario 1, the participants could reach the
destination \emph{Restaurant} by walking a distance similar to that in
other conditions. However, the standard deviation ($35.81$m) is higher
than the standard deviations ($11.45$m and $11.95$m) of the other
conditions, showing that there are larger variations in performances,
due to different paths chosen as shown in
Figure~\ref{fig:result:userstudy1}. In Scenario 2, the relative
difference in standard deviation is even more pronounced ($55.27$m
under the mini-map condition, versus $6.11$m and $7.38$m under the
other conditions), due to the larger differences in walking distances
of the paths chosen. In average, the participants walked a shorter
distance to reach the destination under the full signs or refined
signs conditions ($471.37$m and $476.32$m) than under the mini-map
condition ($503.32$m).

Under the full signs and refined signs conditions, all participants
can reach their destinations. The means and standard deviations of the
walked distances are similar. This shows that the refined sign
placement is as effective as the full sign placement in guiding the
participants to their destinations. However the refined sign placement
uses significantly fewer signs ($3$ signs under refined sign placement
versus $8$ signs under full sign placement in Scenario 1; and $3$
signs under refined sign placement versus $7$ signs under full sign
placement in Scenario 2). Please refer to our supplemental material
for the user study results and a video showing example sessions.

\section{Summary}
\label{sec:summary}
We verify in our experiments that our approach can be applied to
automatically generate wayfinding designs for a variety of layouts,
and that the designs can be used by human users to navigate to their
destinations effectively in virtual worlds. Compared to the
conventional approach of creating wayfinding designs manually, the
novelty of our approach lies in formulating the problem as an
optimization, which can be solved automatically and efficiently, hence
overcoming the design challenge posed by the considerations of
multiple paths and design criteria. Our optimization approach also
allows the flexibility of considering additional constraints in
wayfinding design and the designer can trade off between different
criteria by controlling their corresponding weights. We adopt an
agent-based approach to automatically place signs at strategic
locations, considering human perception and navigation properties such
as eyesight and the possibilities of making mistakes. The agent model
makes it intuitive and flexible for designers to define agent properties
and behaviors according to the specific requirements of their design
projects on hand; signs will be automatically placed according to the
specified agent properties.

\subsection{Limitations and Future Work}
\label{sec:limitation}

Our approach only focuses on placing textual and arrow signs to
facilitate wayfinding. While these are common wayfinding aids, in
reality humans also make use of other wayfinding aids and cues such as
maps (\eg, ``You-are-here'' maps~\cite{handbook,swd}), landmarks and
flow of people movement to determine directions. In future extension
it would be useful to consider all these alternative aids and cues in
generating a wayfinding design. 

% There is previous work in comparing the effectiveness of maps and
% signs for wayfinding~\cite{cliburn2008showing}. One direction for
% future work is to automatically generate both signs and
% maps~\cite{kopf,birsak2014automatic,grabler} for a given virtual
% environment, and allow the user to use both for wayfinding similarly
% as in real-world scenarios. Such a combined approach will likely
% facilitate the level design process and enhance the navigation
% experience offered by the virtual environment.

Our agent-based simulation model only focuses on a few properties that
are relevant to wayfinding. More realistic virtual humans comprising
of cognitive, perceptive, behavioral and kinematic modules, similar to
the autonomous agents used for artificial life
simulation~\cite{shao2005autonomous}, could be used to replace our
agents. The perceptual data obtained from the simulations based on
such agents could be used for more sophisticated wayfinding analysis
to enhance the computationally-generated wayfinding design.

In our current approach, for simplicity we only consider one path for
each source-destination pair. In fact, there could exist multiple
paths (secondary paths) for each pair. This can be modeled by
extending our framework to allow multiple paths for each pair, which
will be considered jointly in the optimization. 

% \blue{Our stochastic optimization only generates an approximation of the
% global optimum. 
% %Figure~\ref{fig:limitation} shows an example. 
% Our optimized solution might deviate slightly from the global optimum found by
% an exhaustive search.}

In our approach, the source-destination pairs are manually specified
rather than automatically generated. This is because our approach does
not infer the layout context. An interesting future direction is to
devise a data-driven approach to automatically identify the possible
locations of interests given a layout based on prior statistics of
human flows, and hence automatically suggest the source-destination
pairs to consider. For example, given a subway station, a data-driven
approach may automatically suggest that (\emph{Entrance}, \emph{Ticket
  Machine}) and (\emph{Ticket Machine}, \emph{Gate}) as likely
source-destination pairs, based on the real-world statistics of human
flows in subway stations.

\subsection*{Acknowledgements}
We thank Ana Aravena for narrating the demonstration video. This
research is supported by the UMass Boston StartUp Grant
P20150000029280 and by the Joseph P. Healey Research Grant Program
provided by the Office of the Vice Provost for Research and Strategic
Initiatives \& Dean of Graduate Studies of UMass Boston. This research
is also supported by the National Science Foundation under award
number 1565978. We acknowledge NVIDIA Corporation for graphics card
donation.

\bibliographystyle{IEEEtran}
\bibliography{vr_attention}

%\newpage
\begin{IEEEbiography}[{\includegraphics[width=1in,height=1.25in,clip,keepaspectratio]
   {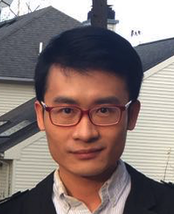}}]
 {Haikun Huang} is a PhD student at the University of Massachusetts
 Boston. He received his BSc degree in computer science from the
 University of Massachusetts Boston in 2016. His research interests
 include computer graphics and visualization. He is a member of the
 IEEE.
\end{IEEEbiography}
\vspace{-5mm}

\begin{IEEEbiography}[{\includegraphics[width=1in,height=1.25in,clip,keepaspectratio]
   {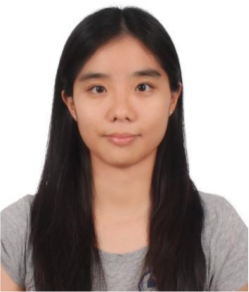}}]
 {Ni-Ching (Monica) Lin} is a graduate student in the Institute of
 Electrical and Control Engineering, National Chiao Tung University,
 Taiwan. She received her BEng degrees with honors in Electrical
 Engineering from Tamkang University, Taiwan. Her research interests
 include robotic vision, 3D visualization, and robot navigation.
\end{IEEEbiography}
\vspace{-5mm}

\begin{IEEEbiography}[{\includegraphics[width=1in,height=1.25in,clip,keepaspectratio]
   {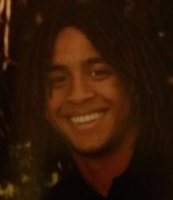}}]
 {Lorenzo Barrett} is an undergraduate student in computer science at
 the University of Massachusetts Boston. His research interests
 include visualization and cyber security.
\end{IEEEbiography}
\vspace{-5mm}

\begin{IEEEbiography}[{\includegraphics[width=1in,height=1.25in,clip,keepaspectratio]
   {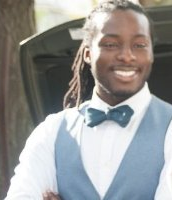}}]
 {Darian Springer} is an undergraduate student in computer science at
 the University of Massachusetts Boston. His research interests
 include computer graphics and visualization.
\end{IEEEbiography}
\vspace{-5mm}

\begin{IEEEbiography}[{\includegraphics[width=1in,height=1.25in,clip,keepaspectratio]
   {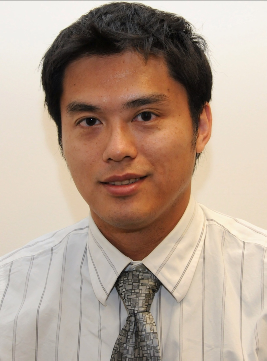}}]
 {Hsueh-Cheng (Nick) Wang} is an assistant professor in the Electrical
 and Computer Engineering and Institute of Electrical and Control
 Engineering at National Chiao Tung University, Taiwan. Dr. Wang and
 his research group focus on developing robotic systems to solve
 real-world problems in direct support of individuals.
\end{IEEEbiography}
\vspace{-5mm}

\begin{IEEEbiography}[{\includegraphics[width=1in,height=1.25in,clip,keepaspectratio]
   {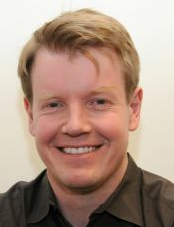}}]
  {Marc Pomplun} is a professor of computer science at the University
  of Massachusetts at Boston and the Director of the Visual
  Attention Laboratory. His work focuses on analysing, modelling and
  simulating aspects of human vision.
\end{IEEEbiography}
\vspace{-5mm}

\begin{IEEEbiography}[{\includegraphics[width=1in,height=1.25in,clip,keepaspectratio]
   {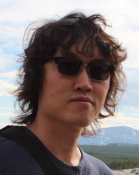}}]
 {Lap-Fai (Craig) Yu} is an assistant professor at the University of
 Massachusetts Boston, where he directs the Graphics and Virtual
 Environment Laboratory. He received his BEng and MPhil degrees in
 computer science from the Hong Kong University of Science and
 Technology (HKUST) in 2007 and 2009 respectively, and his PhD degree
 in computer science from the University of California, Los Angeles,
 in 2013. He was a visiting scholar at Stanford University and a
 visiting scientist at the Massachusetts Institute of Technology. His
 research interests include computer graphics and computer vision. He
 served in the program committee of Pacific Graphics 2016 and 2017,
 and the ACM SIGGRAPH Symposium on Interactive 3D Graphics and Games
 (i3D) 2016. He is a member of the IEEE.
\end{IEEEbiography}

\end{document}